\documentclass[onecolumn,superscriptaddress]{revtex4-1}
\usepackage{bm, mathrsfs, amsmath, amssymb}
\usepackage[dvipdfmx]{graphicx}

\begin{document}

\title{
Rapidity window dependences of higher order cumulants
and diffusion master equation
}

\author{
Masakiyo Kitazawa
}
\affiliation{Department of Physics, Osaka University, Toyonaka,
Osaka 560-0043, Japan}

\date{\today}


\begin{abstract}

We study the rapidity window dependences of higher order 
cumulants of conserved charges observed in relativistic
heavy ion collisions.
The time evolution and the rapidity window dependence of 
the non-Gaussian fluctuations are described by the diffusion 
master equation.
Analytic formulas for the time evolution of cumulants in a 
rapidity window are obtained for arbitrary initial conditions.
We discuss that the rapidity window dependences of the 
non-Gaussian cumulants have characteristic structures 
reflecting the non-equilibrium property of fluctuations, 
which can be observed in relativistic heavy ion collisions 
with the present detectors.
It is argued that various information on the thermal and 
transport properties of the hot medium can be revealed 
experimentally by the study of the rapidity window 
dependences, especially by the combined use, of the higher 
order cumulants.
Formulas of higher order cumulants for a probability distribution 
composed of sub-probabilities, which are useful for various 
studies of non-Gaussian cumulants, are also presented.

\end{abstract}

\maketitle

\section{Introduction}
\label{sec:intro}

Bulk fluctuations of conserved charges are believed to be 
useful observables for the study of thermal property of the 
primordial medium created in relativistic heavy ion collisions.
Active experimental analyses of fluctuation observables have 
been performed in the beam-energy scan (BES) program \cite{BES-II}
at the Relativistic Heavy Ion Collider (RHIC) 
\cite{STAR,PHENIX,Luo:2012kja},
as well as the Large Hadron Collider (LHC) \cite{ALICE}.
These experiments have been observed not only the 
second order but also higher order cumulants
characterizing non-Gaussianity of the fluctuations 
by the event-by-event analysis.
Various theoretical suggestions have been made to utilize 
fluctuation observables as signals of the 
deconfinement transition and the QCD critical point
\cite{Koch:2008ia,Stephanov:1998dy,Asakawa:2000wh,Jeon:2000wg,
Ejiri:2005wq,Stephanov:2008qz,Asakawa:2009aj,HRG,Friman:2011pf,
Stephanov:2011pb,KA1,KA2,Morita:2013tu,KAO,Herold:2014zoa,SAK}.
The cumulants of conserved charges in the 
equilibrated medium are also observable in lattice QCD 
Monte Carlo simulations, and the numerical study 
of various cumulants have been carried out actively
\cite{Gupta:2011wh,Bazavov:2012jq,Bazavov:2012vg,Bazavov:2013dta,
Nakamura:2013ska,Bellwied:2013cta,Borsanyi:2014ewa,Bazavov:2014yba,
Gupta:2014qka}.

One of the most notable features of the measurement of 
fluctuations in heavy ion collisions is that higher order 
cumulants of particle numbers in a rapidity window have been 
observed with moderate statistics up to fourth order.
Such measurements are possible because the systems observed 
in these experiments are not large; the particle number 
measured in one event is at most of order $10^3$ \cite{STAR,ALICE}, 
although the local equilibration is expected to realize
for some period in the time evolution.
Experimental detectors are designed to observe and identify 
almost all charged particles entering the detector.
They thus can provide {\it full counting statistics} \cite{FCS} 
of the event-by-event distribution of the particle numbers 
up to the error arising from the acceptance, efficiency and 
particle miss identifications \cite{KA2,Bzdak:2012ab,Garg:2012nf,
OAK,Nahrgang:2014fza} 
of the detectors.

Because the experimental detectors in heavy ion collisions 
count particle numbers which arrive at the detectors, 
the event-by-event analysis can observe fluctuations 
in the final state of the collision events.
If the fluctuations are fully equilibrated in the final 
state, they can only tell us information on the thermodynamics 
at this time, which is in the hadronic medium and well 
described by the hadron resonance gas (HRG) model \cite{HRG}.
The experimental results, however, suggest that the observed 
fluctuations are not the one of the equilibrated hadronic medium.
First, the experimental results on the cumulants and their 
ratios have statistically-significant deviation from 
the values in the HRG model \cite{ALICE,STAR}.
In the HRG model, the fluctuations of net particle numbers,
including net-baryon and net-electric charge which are conserved 
charges, are given by the Skellam distribution to a good 
approximation \cite{HRG}.
The cumulants observed at RHIC and LHC show deviations 
from these Skellam values.
The second observation is concerned with the rapidity 
window dependences of the cumulants.
In a thermal medium in grand canonical ensemble, 
cumulants of conserved charges are extensive variables and 
proportional to volume.
In heavy ion collisions, fluctuations of particle number 
in a given pseudo-rapidity window, $\Delta \eta$, are observed.
Assuming the Bjorken expansion and that the pseudo-rapidity window 
can be used as a proxy of the one of the coordinate-space rapidity,
$\Delta y$ \cite{OAKS}, the spatial volume to count the particle 
number is proportional to $\Delta \eta$.
The cumulants, therefore, should be proportional 
to $\Delta \eta$ if the fluctuations are fully equilibrated.
The net-electric charge fluctuation 
$\langle (\delta N_{\rm Q})^2 \rangle$ 
observed by ALICE collaboration \cite{ALICE}, however, 
clearly shows that this proportionality is violated.
A similar violation is also indicated in the higher order 
cumulants of net-proton number observed by STAR collaboration 
\cite{Luo:2012kja}.
It is this non-equilibrium nature of fluctuations which enables 
us to explore the primordial thermodynamics using fluctuation
observables \cite{Asakawa:2000wh,Jeon:2000wg}.

For conserved charges, the off-equilibrium property of 
fluctuations in the final state is not surprising, because 
the modification of local density of a conserved charge is 
proceeded only by the diffusive processes, which is typically 
slow; because of the conservation law, the time scale of the 
diffusive process can become arbitrary slow as the spatial 
scale becomes larger.
In fact, earlier studies suggested that the time scale of the 
diffusive process for conserved charges is so slow 
that the thermal fluctuations generated 
in the deconfined medium survive until the final state 
when $\Delta \eta$ is taken sufficiently large 
\cite{Asakawa:2000wh,Jeon:2000wg,Shuryak:2000pd,Aziz:2004qu}.
Because the effect of the diffusion with a fixed $\Delta \eta$ is 
determined by the transport of the charges, the conservation law 
also suggests that one can investigate the transport property of 
the hot medium, such as the magnitude of the diffusion constant, 
experimentally using the $\Delta \eta$ dependence of 
conserved-charge fluctuations \cite{Shuryak:2000pd,ALICE,KAO}.

In Refs.~\cite{Asakawa:2000wh,Jeon:2000wg,Shuryak:2000pd,Aziz:2004qu}, 
the time evolution of fluctuations are considered 
only for Gaussian fluctuations.
Because the heavy ion collisions can measure 
the higher order cumulants, 
the $\Delta \eta$ dependences of not only Gaussian but also 
non-Gaussian fluctuations should encode more information 
on the thermal and transport properties of the medium \cite{KAO}.
To extract such information, however, an appropriate description 
of the non-equilibrium time evolution of the non-Gaussian 
fluctuations is required.

When one investigates the diffusive process of non-Gaussian 
fluctuations in heavy ion collisions, it is important to keep 
in mind that the cumulants observed so far are not far from 
the ones of the Skellam distribution \cite{STAR,ALICE}, 
which are the equilibrated values in the HRG model.
A possible interpretation of these results is that 
the signals in cumulants are developed in the primordial stage,
but they are blurred by the diffusive process in the hadronic stage.
In fact, the experimental result in Ref.~\cite{ALICE}
indicates that the fluctuations {\it increase} toward
the equilibrated value in the hadronic medium 
due to the diffusion \cite{Shuryak:2000pd,KAO,SAK}.
If this is the case, it is important to describe the approach
of the cumulants toward the hadronic value in the analysis of 
the diffusive process in the hadronic medium.
For this purpose, the model for the diffusive process should 
have nonzero higher order cumulants in equilibrium 
which are consistent with those in the hadronic medium.

In Ref.~\cite{KAO}, to describe the time evolution of 
non-Gaussian fluctuations 
the diffusion master equation (DME) is employed.
As discussed in Ref.~\cite{KAO}, in this model the fluctuations 
in equilibrium are given by the Poisson or Skellam distribution 
as a consequence of the discrete nature of the particle number.
The time evolutions of the second and fourth order cumulants 
of net particle numbers are studied in this model.
It is found that the fourth-order cumulant can show 
characteristic behaviors as a function of $\Delta\eta$
and tends to be suppressed compared with the second order one,
if the thermal fluctuations in the primordial medium are suppressed.
These results can be confirmed experimentally with the 
present detectors.
The analysis is later extended to finite volume systems \cite{SAK}
to investigate the effect of the global charge conservation.

The purpose of the present paper is to elaborate on 
detailed discussions skipped in Ref.~\cite{KAO}.
In the present study, we show the analytic 
procedure for obtaining the time evolution of cumulants 
in the DME with arbitrary initial conditions.
We also extend the discussion on the $\Delta\eta$ dependence
of cumulants in Ref.~\cite{KAO} to more general cases.
The $\Delta\eta$ dependence of the third order cumulant is
considered in addition to second and fourth order ones 
discussed in Ref.~\cite{KAO}.
The initial condition is also extended to more general cases.
Using these results, we pursue the possibility to extract
various information on the thermal and transport properties
of the hot medium from the non-Gaussian cumulants of 
net-baryon number and net-electric charge in heavy ion 
collisions.

This paper is organized as follows.
In Sec.~\ref{sec:DME}, we present the analytic solution 
for the time evolution of cumulants in the DME.
The result is then extended in Sec.~\ref{sec:multi} to the 
multi-particle systems in order to describe the cumulants of 
net particle numbers.
In Sec.~\ref{sec:Dy}, we apply these results
to the description of $\Delta\eta$ dependences of non-Gaussian
cumulants in heavy ion collisions.
Some discussions and a short summary is given in Sec.~\ref{sec:summary}.
In Appendix~\ref{app:superposition}, we derive formulas
of cumulants for a probability given by a superposition of 
sub-probabilities, which are used in Sec.~\ref{sec:DME}
to obtain the solution of the DME for arbitrary initial 
conditions.
In Appendix~\ref{app:chemical}, we consider a simple 
stochastic model for a chemical reaction.
This appendix will help readers who are not familiar with 
stochastic models like the DME to understand Sec.~\ref{sec:DME}.

\section{Diffusion master equation}
\label{sec:DME}

To describe the diffusion of non-Gaussian 
fluctuations of the conserved charges in heavy ion collisions,
we use the diffusion master equation (DME) \cite{KAO}.
After introducing the DME and describing its general 
property in Sec.~\ref{sec:DMEintro}, we derive the analytic form of 
the time evolution of the cumulants for fixed initial conditions
in Secs.~\ref{sec:fixed} - \ref{sec:uniform}.
The solution is then extended to general initial conditions
in Sec.~\ref{sec:general}.

\subsection{Diffusion master equation}
\label{sec:DMEintro}

\begin{figure}
\begin{center}
\includegraphics[width=.49\textwidth]{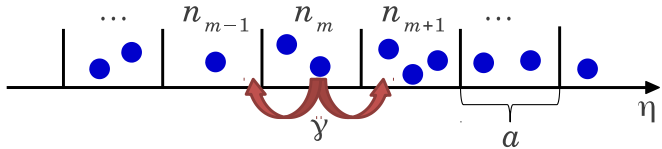}
\caption{
System described by the diffusion master 
equation (DME) Eq.~(\ref{eq:DME}).
}
\label{fig:DME}
\end{center}
\end{figure}

We consider a single species of classical Brownian 
particles moving in one dimensional space.
To describe microscopic states of this system, 
we divide the spatial coordinate into 
discrete cells with an equal length $a$, as in 
Fig.~\ref{fig:DME}.
We label each cell by an integer $m$ and denote 
the number of particles in each cell by $n_m$.
A microscopic state is then specified by 
the vector $\bm{n}=(\cdots,n_{m-1},n_m,n_{m+1},\cdots)$ 
representing particle numbers in all cells.
One can also introduce the probability distribution function 
$P(\bm{n},t)$ that each cell contains $n_m$ particles at time $t$.

Next, we assume that each particle moves adjacent cells
with a probability $\gamma(t)$ per unit time.
The probability $P(\bm{n},t)$ then follows the 
first-order differential equation \cite{Gardiner}
\begin{align}
\frac{\partial}{\partial t}P(\bm{n},t)
= \gamma(t) \sum_m & [
( n_m + 1 ) \{ P(\bm{n}+\bm{e}_{m+1}-\bm{e}_{m-1})
+ P(\bm{n}-\bm{e}_{m+1}+\bm{e}_{m-1}) \}
\nonumber \\
& -2 n_m P(\bm{n},t) ],
\label{eq:DME}
\end{align}
where $\bm{e}_m = (0,\cdots,0,1,0,\cdots,0)$ is the vector 
that all components are 
zero except for the $m$th-one which takes unity.
After solving Eq.~(\ref{eq:DME}), we 
take the continuum limit defined by $a\to0$
to recover the continuity of the spatial coordinate.
Although the generalization of the model to a multi-dimensional
system is straightforward, in this study we limit our attention 
to the one-dimensional problem to avoid unnecessary complexity.
Note that the one-dimensional model is sufficient 
for the description of the diffusive process along 
the rapidity direction in heavy ion collisions.

As we will see later, 
after taking the continuum limit 
the time evolution of the average particle number density 
$n(x,t)$ (the exact definition will be given 
in Sec.~\ref{sec:continuum}) in the DME 
obeys the standard diffusion equation 
\begin{align}
\frac{\partial}{\partial t} n(x,t)
= D(t) \frac{\partial^2}{\partial x^2} n(x,t),
\label{eq:diffusion}
\end{align}
with the diffusion constant $D(t) = \gamma(t) a^2$.
It is also shown \cite{KAO} that the time evolution of the 
Gaussian fluctuations of $n(x,t)$ in this limit is consistent 
with the {\it stochastic diffusion equation}
\begin{align}
\frac{\partial}{\partial t} n(x,t)
= \frac{\partial}{\partial x} 
\left( D(t) \frac{\partial}{\partial x} n(x,t) + \xi(x,t) \right),
\label{eq:SDE}
\end{align}
where $\xi(x,t)$ represents the stochastic term.
Here, it is assumed that $\xi(x,t)$ is local with respect to 
$x$ and $t$, i.e. $\langle \xi(x_1,t_1) \xi(x_2,t_2) \rangle$ 
is proportional to $\delta(x_1-x_2)\delta(t_1-t_2)$. 
In this case the property of $\xi(x,t)$ is completely 
determined by the fluctuation dissipation relation \cite{Landau}.

The DME, on the other hand, has a property that the 
higher order cumulants take nonzero values in equilibrium 
defined by $t\to\infty$ limit.
This property is contrasted with Eq.~(\ref{eq:SDE}) in 
which the fluctuation of $n(x,t)$ becomes Gaussian 
and all the cumulants of $n(x,t)$ higher than the second order
vanish in equilibrium for $n$ independent $D(t)$ \cite{KAO}.
The higher order cumulants of the 
particle number in some range of the spatial coordinate
in the DME in equilibrium are given by the Poissonian ones
\cite{KAO}.
This property is easily understood from the fact that 
in equilibrium each Brownian particle exists any place with an 
equal probability and the individual particles are 
uncorrelated.
When the model is extended to treat the net-particle number
in Sec.~\ref{sec:multi},
the distribution is given by the Skellam one.
This property is suitable to model the time evolution
of higher order cumulants in heavy ion collisions.

\subsection{Factorial generating function for fixed initial condition}
\label{sec:fixed}

In the following, we solve the DME Eq.~(\ref{eq:DME}).
For readers who are not familiar with stochastic 
equations like Eq.~(\ref{eq:DME}), we recommend them 
to read Appendix~\ref{app:chemical} before this section.

In the following five subsections, 
we first solve Eq.~(\ref{eq:DME}) for the fixed initial 
condition,
\begin{align}
P(\bm{n},0) = \prod_m \delta_{n_m,M_m} ,
\label{eq:P0fixed}
\end{align}
and present $t$ dependence of cumulants of $P(\bm{n},t)$.
In Eq.~(\ref{eq:P0fixed}), the particle number in each 
cell, $n_m$, are fixed to $M_m$ without fluctuations at $t=0$.
The analysis is later generalized to initial conditions 
having fluctuations in Sec.~\ref{sec:general}.

To solve Eq.~(\ref{eq:DME}) with the initial condition
Eq.~(\ref{eq:P0fixed}), it is convenient to 
use the factorial generating function \cite{Gardiner}
\begin{align}
G_{\rm f} ( \bm{s},t )
= \sum_{\bm{n}} \bigg( \prod_m s_m^{n_m} \bigg) P( \bm{n},t ),
\label{eq:G_f}
\end{align}
where the sum is taken over all possible combinations of $n_m$.
Substituting Eq.~(\ref{eq:G_f}) in Eq.~(\ref{eq:DME}),
one obtains 
\begin{align}
\frac{\partial}{\partial t} G_{\rm f} (\bm{s},t)
= \gamma(t) \sum_m ( s_{m+1} - 2s_m + s_{m-1} )
\frac{\partial}{\partial s_m} G_{\rm f} (\bm{s},t) .
\label{eq:G_f:ME}
\end{align}
Because Eq.~(\ref{eq:G_f:ME}) is a first-order partial 
differential equation, this equation is solved by 
the method of characteristics.
In this method, we use the fact that 
the solution of Eq.~(\ref{eq:G_f:ME}) satisfies 
\begin{align}
\frac{d}{dt} G_{\rm f}( \bm{s}^c(t),t)=0,
\label{eq:G=0}
\end{align}
where the characteristic line $s^c_m(t)$ is the 
solution of the characteristic equations
\begin{align}
\frac{ds^c_m}{dt} = -\gamma(t) ( s^c_{m+1} - 2s^c_m + s^c_{m-1} ).
\label{eq:char}
\end{align}

Equation~(\ref{eq:char}) is easily solved in Fourier space.
Assuming the periodic boundary condition and defining 
\begin{align}
r_j = \frac1N \sum_m s_m e^{2\pi i jm/N},
\label{eq:FT:ts}
\end{align}
with $N$ being the total number of cells, 
the characteristic equations become 
\begin{align}
\frac{ d}{dt} r^c_j 
= -\gamma(t) ( e^{2\pi i j/N} + e^{-2\pi i j/N} -2 ) r^c_j
\equiv \omega_j(t) r^c_j,
\label{eq:dr^c}
\end{align}
with 
\begin{align}
\omega_j(t) = -\gamma(t) ( e^{2\pi i j/N} + e^{-2\pi i j/N} -2 )
\simeq \gamma(t) \left( \frac{2\pi}N j \right)^2,
\end{align}
where the last nearly-equality is satisfied for $j/N\ll1$.
The solution of Eq.~(\ref{eq:dr^c}) is
\begin{align}
r^c_j(t) = r^c_j(0) \exp[ \int_0^t dt' \omega_j(t')].
\label{eq:r^c_j}
\end{align}
When $\gamma(t)$ does not have $t$ dependence, 
Eq.~(\ref{eq:r^c_j}) has a simple form
\begin{align}
r^c_j(t) = r^c_j(0) e^{\omega_j t}.
\end{align}

The initial condition Eq.~(\ref{eq:P0fixed})
is transferred to the factorial generating function
$G_{\rm f}(\bm{s},t)$ as 
\begin{align}
G_{\rm f}(\bm{s},0) = \prod_m s_m^{M_m} 
= \prod_m \left( \sum_j r_j(0) e^{2\pi i jm/N} \right)^{M_m}.
\label{eq:G0fixed}
\end{align}
Because of Eq.~(\ref{eq:G=0}), the solution of 
Eq.~(\ref{eq:G_f:ME}) is obtained by substituting 
Eq.~(\ref{eq:r^c_j}) in Eq.~(\ref{eq:G0fixed}) 
which yields
\begin{align}
G_{\rm f} ( \bm{s}, t )
= \prod_m \left( \sum_j r_j e^{2\pi i jm/N} \exp[ -\int_0^t dt' \omega_j(t')] \right)^{M_m}.
\label{eq:G_f(t)}
\end{align}
The corresponding factorial cumulant generating function is given by
\begin{align}
K_{\rm f} ( \bm{s},t ) = \log G_{\rm f} ( \bm{s}, t )
= \sum_m M_m \log \sum_j r_j e^{2\pi i jm/N} \exp[ -\int_0^t dt' \omega_j(t')] .
\label{eq:K_f(t)}
\end{align}

\subsection{Cumulants and factorial cumulants}
\label{sec:cumu}

Using Eq.~(\ref{eq:K_f(t)}), the factorial cumulants of 
$n_m$ are given by
\begin{align}
\langle n_{m_1} n_{m_2} \cdots n_{m_l} \rangle_{\rm fc}
=& \left. \frac{\partial^l  K_{\rm f} }
{\partial s_{m_1} \partial s_{m_2} \cdots \partial s_{m_l}}
\right|_{\bm{s}=\bm{1}},
\end{align}
where $\bm{s}=\bm{1}$ means that $s_m=1$ for all $m$.
Note that this condition corresponds to $r_0=1$ 
and $r_j=0$ for $j\ne0$.
Using this relation, factorial cumulants in the Fourier space 
are calculated to be
\begin{align}
\langle \tilde{n}_{k_1} \tilde{n}_{k_2} \cdots \tilde{n}_{k_l} \rangle_{\rm fc}
=& 
\sum_{m_1,\cdots,m_l}
e^{-2\pi i (k_1 m_1 + \cdots k_l m_l ) /N}
\langle \tilde{n}_{m_1} \cdots \tilde{n}_{m_l} \rangle_{\rm fc}
\nonumber \\
=&
\sum_{m_1,\cdots,m_l}
e^{-2\pi i (k_1 m_1 + \cdots k_l m_l ) /N}
\left. \frac{\partial^l  K_{\rm f} }
{\partial s_{m_1} \cdots \partial s_{m_l}}
\right|_{\bm{s}=\bm{1}}
\nonumber \\
=&
\left. \frac{\partial^l  K_{\rm f} }
{\partial r_{k_1} \cdots \partial r_{k_l}}
\right|_{\bm{s}=\bm{1}},
\label{eq:<nk>_fc}
\end{align}
where the Fourier transform of the particle number
 is defined by
\begin{align}
\tilde{n}_k = \sum_x n_m e^{-2\pi i km/N},
\quad
n_m = \frac1N \sum_k \tilde{n}_k e^{2\pi i km/N}.
\end{align}
Substituting Eq.~(\ref{eq:K_f(t)}) in Eq.~(\ref{eq:<nk>_fc}),
one obtains the factorial cumulants up to fourth order as 
\begin{align}
\langle \tilde{n}_k \rangle_{\rm fc}
=& \tilde{M}_k 
\exp[ -\int_0^t dt' \omega_k(t')],
\label{eq:nk_fc1} \\
\langle \tilde{n}_{k_1} \tilde{n}_{k_2} \rangle_{\rm fc}
=& - \tilde{M}_{k_1+k_2} 
\exp[ -\int_0^t dt' ( \omega_{k_1}(t') + \omega_{k_2}(t') ) ],
\label{eq:nk_fc2} \\
\langle \tilde{n}_{k_1} \tilde{n}_{k_2} \tilde{n}_{k_3} \rangle_{\rm fc}
=& 2 \tilde{M}_{k_1+k_2+k_3} 
\exp[ -\int_0^t dt' ( \omega_{k_1}(t') + \omega_{k_2}(t') + \omega_{k_3}(t') ) ],
\label{eq:nk_fc3} \\
\langle \tilde{n}_{k_1} \tilde{n}_{k_2} \tilde{n}_{k_3} 
\tilde{n}_{k_4} \rangle_{\rm fc}
=& -6 \tilde{M}_{k_1+k_2+k_3+k_4} 
\exp[ -\int_0^t dt' ( \omega_{k_1}(t') + \omega_{k_2}(t')
 + \omega_{k_3}(t') + \omega_{k_4}(t') ) ],
\label{eq:nk_fc4}
\end{align}
with $\tilde{M}_k = \sum_m M_m e^{-2\pi ikm/N}$.
Although we do not show the explicit forms of cumulants higher 
than fourth order, they can be obtained straightforwardly with 
a similar manipulation.

Cumulants of $\tilde{n}_k$ are obtained by using 
cumulant generating function
\begin{align}
K(\bm{\theta},t) = K_{\rm f}(\bm{s},t)|_{s_i=e^{\theta_i}},
\end{align}
as
\begin{align}
\langle n_{m_1} n_{m_2} \cdots n_{m_l} \rangle_{\rm c}
=& \left. \frac{\partial^l  K }
{ \partial \theta_1 \cdots \partial \theta_l } \right|_{\bm{\theta}=0} .
\end{align}
Using the relations 
\begin{align}
\frac{\partial s_{m}}{\partial \theta_{m'}} = \delta_{m, m'}, \quad
\frac{\partial^2 s_{m}}{\partial \theta_{m_1} \partial \theta_{m_2}} 
= \delta_{m, {m_1}} \delta_{m, {m_2}},
\end{align}
and so forth, one finds that the cumulants of $\tilde{n}_k$ 
are related to factorial cumulants Eqs.~(\ref{eq:nk_fc1}) - 
(\ref{eq:nk_fc4}) as 
\begin{align}
\langle \tilde{n}_k \rangle_{\rm c}
=& \langle \tilde{n}_k \rangle_{\rm fc} ,
\\
\langle \tilde{n}_{k_1} \tilde{n}_{k_2} \rangle_{\rm c}
=& \langle \tilde{n}_{k_1} \tilde{n}_{k_2} \rangle_{\rm fc}
+ \langle \tilde{n}_{k_1+k_2} \rangle_{\rm fc} ,
\\
\langle \tilde{n}_{k_1} \tilde{n}_{k_2} \tilde{n}_{k_3} \rangle_{\rm c}
=& \langle \tilde{n}_{k_1} \tilde{n}_{k_2} \tilde{n}_{k_3} \rangle_{\rm fc}
+ \langle \tilde{n}_{k_1} \tilde{n}_{k_2+k_3} \rangle_{\rm fc} + {\rm (comb.)}
\nonumber \\
&+ \langle \tilde{n}_{k_1+k_2+k_3} \rangle_{\rm fc} ,
\\
\langle \tilde{n}_{k_1} \tilde{n}_{k_2} \tilde{n}_{k_3} \tilde{n}_{k_4} \rangle_{\rm c}
=& 
\langle \tilde{n}_{k_1} \tilde{n}_{k_2} \tilde{n}_{k_3} \tilde{n}_{k_4} \rangle_{\rm fc}
+ \langle \tilde{n}_{k_1} \tilde{n}_{k_2} \tilde{n}_{k_3+k_4} \rangle_{\rm fc} + {\rm (comb.)}
\nonumber \\
&+ \langle \tilde{n}_{k_1+k_2} \tilde{n}_{k_3+k_4} \rangle_{\rm fc} + {\rm (comb.)}
+ \langle \tilde{n}_{k_1} \tilde{n}_{k_2+k_3+k_4} \rangle_{\rm fc} + {\rm (comb.)}
\nonumber \\
&+ \langle \tilde{n}_{k_1+k_2+k_3+k_4} \rangle_{\rm fc} ,
\end{align}
where (comb.) means the sum for all possible combinations
for subscripts of $\tilde{n}_k$; for example,
\begin{align}
\langle \tilde{n}_{k_1} \tilde{n}_{k_2+k_3} \rangle_{\rm fc} + {\rm (comb.)}
= \langle \tilde{n}_{k_1} \tilde{n}_{k_2+k_3} \rangle_{\rm fc}
+ \langle \tilde{n}_{k_2} \tilde{n}_{k_3+k_1} \rangle_{\rm fc}
+ \langle \tilde{n}_{k_3} \tilde{n}_{k_1+k_2} \rangle_{\rm fc}.
\end{align}

\subsection{Continuum limit}
\label{sec:continuum}

Next, we take the continuum limit, $a\to0$.
First, we denote the spatial coordinate of the $m$th cell as 
$x=ma$.
In the continuum limit, using the particle number density 
defined by $n(x) = n_m/a$, the probability distribution 
function in Eq.~(\ref{eq:DME}) is promoted to the functional 
$P[n(x),t]$ of $n(x)$.
Note, however, that this notation is conceptual;
in actual applications, the functional $P[n(\eta),\tau]$ 
is always understood as the limit of the function $P(\bm{n},\tau)$ 
with small but finite $a$.
The Fourier transform of $n(x)$ is 
$\tilde{n}(p) = \int dx e^{-ipx} n(x)$ with $p = 2\pi k / Na$.
One can easily verify that the factorial cumulants of 
$\tilde{n}(p)$ in the continuum notation are 
obtained by replacing $\tilde{n}_k \to \tilde{n}(p)$ 
in Eqs.~(\ref{eq:nk_fc1}) - (\ref{eq:nk_fc4}) with
\begin{align}
\omega_p(t) = \gamma(t) a^2 p^2.
\end{align}

We require that the deterministic part of Eq.~(\ref{eq:DME}) 
obeys the solution of the diffusion equation
Eq.~(\ref{eq:diffusion}) in the continuum limit.
The solution of Eq.~(\ref{eq:diffusion}) in Fourier space 
with the initial condition $n(x,0)=M(x)$ is 
\begin{align}
\tilde{n}(p,t) 
= \tilde{M}(p) \exp[-\int_0^t dt' D(t')p^2 ].
\label{eq:diffusion:solution}
\end{align}
By comparing Eq.~(\ref{eq:diffusion:solution}) 
with Eq.~(\ref{eq:nk_fc1}), one finds that 
these equations give the same result when one takes
\begin{align}
D(t) = \gamma(t) a^2.
\label{eq:D(t)}
\end{align}
The $a\to0$ limit therefore has to be taken with 
Eq.~(\ref{eq:D(t)}).
Similarly, one can show that the time evolution of 
the second order cumulant in the DME, Eq.~(\ref{eq:nk_fc2}), 
in this limit agrees with the solution of the stochastic diffusion 
equation Eq.~(\ref{eq:SDE}) \cite{KAO}.

\subsection{Particle number in a volume}
\label{sec:volume}

Next, we investigate the cumulants 
of the particle number in a range with length $\Delta$,
\begin{align}
Q(\Delta,t) = \int_{-\Delta/2}^{\Delta/2} dx n(x,t).
\end{align}
To represent the cumulants of $Q$,
we first note that the diffusion distance of each Brownian 
particle in the model is given by 
\begin{align}
d(t) = \sqrt{ 2 \int_0^t dt' D(t') },
\end{align}
from Eq.~(\ref{eq:diffusion:solution})
\cite{Einstein,SAK}.
Using this quantity having the dimension of length,
one can define a dimensionless variable
\begin{align}
X = \frac{d(t)}\Delta.
\label{eq:X}
\end{align}
The $t$ and $\Delta$ dependences of 
$\langle (Q(\Delta,t))^n \rangle_c/\Delta$ 
then are described only through this dimensionless variable,
because $\langle (Q(\Delta,t))^n \rangle_c/\Delta$ is dimensionless 
and Eq.~(\ref{eq:X}) is the natural dimensionless variable which 
can be created from the combination of $\Delta$, $t$ and $D(t)$.

Dependences of the first order cumulant on $\Delta$ and $t$ 
is calculated to be 
\begin{align}
\langle Q(\Delta,t) \rangle_c 
=& \int_{-\Delta/2}^{\Delta/2} dx \langle n(x) \rangle_c
= \int_{-\Delta/2}^{\Delta/2} dx \int \frac{dp}{2\pi} e^{ipx}
\langle \tilde{n}(p) \rangle_c
\nonumber \\
=& \int_{-\Delta/2}^{\Delta/2} dx \int \frac{dp}{2\pi} e^{ipx}
\tilde{M}(p) e^{-\omega_pt}
\nonumber \\
=& \int dz M(z) \int_{-\Delta/2}^{\Delta/2} dx \int \frac{dp}{2\pi} 
e^{ip(x-z)} e^{-Dtp^2}
\nonumber \\
=& \int dz M( z ) I_X(z/\Delta).
\label{eq:<Q^1>}
\end{align}
Similarly, the second order cumulant of $Q(\Delta,t)$ is calculated to be
\begin{align}
\langle (Q(\Delta,t))^2 \rangle_c 
=& \int_{-\Delta/2}^{\Delta/2} dx_1 dx_2 
\int \frac{dp_1 dp_2}{(2\pi)^2} e^{i( p_1 x_1 + p_2 x_2 )}
\langle \tilde{n}(p_1) \tilde{n}(p_2) \rangle_c
\nonumber \\
=& \int_{-\Delta/2}^{\Delta/2} dx_1 dx_2 
\int \frac{dp_1 dp_2}{(2\pi)^2} e^{i( p_1 x_1 + p_2 x_2 )}
\nonumber \\
& \times
\tilde{M}(p_1+p_2) ( e^{-\omega_{p_1+p_2}t} - e^{-(\omega_{p_1}+\omega_{p_2})t} )
\nonumber \\
=& \int dz M(z) 
\int_{-\Delta/2}^{\Delta/2} dx_1 dx_2 
\int \frac{dp_1 dp_2}{(2\pi)^2} e^{i( p_1 (x_1-z) + p_2 (x_2-z) )}
\nonumber \\
& \times ( e^{-Dt(p_1+p_2)^2} - e^{-Dt( p_1^2 +p_2^2 ) } )
\nonumber \\
=&  \int dz M( z ) 
\left( I_X(z/\Delta) - I_X(z/\Delta)^2 \right).
\label{eq:<Q^2>}
\end{align}
To obtain the last line, it is convenient to use a relation
\begin{align}
\int \frac{ dp_1 dp_2 }{ (2\pi)^2 }
e^{ ip_1 x_1 } e^{ ip_2 x_2 } f(p_1+p_2)
= \delta( x_1-x_2 ) \int \frac{dp}{2\pi}
e^{ipx_1} f(p) .
\end{align}
In the last line we have defined 
\begin{align}
I_X(\zeta) = \int_{-1/2}^{1/2} d\xi \int \frac{dp}{2\pi}
e^{-X^2 p^2/2} e^{ip(\xi+\zeta)}
= \frac12 \left(
{\rm erf} \frac{ \zeta+1/2 }{\sqrt2 X} 
- {\rm erf} \frac{ \zeta-1/2 }{\sqrt2 X} \right),
\label{eq:I_X}
\end{align}
where ${\rm erf}(x)= (2/\sqrt{\pi})\int_0^x dt e^{-t^2}$ 
is the error function.
Some properties of Eq.~(\ref{eq:I_X}) is summarized 
in appendix~\ref{app:I_X}.

Repeating similar manipulations, 
one finds that higher order cumulants of $Q(\Delta,t)$ 
up to fourth order are given by 
\begin{align}
\langle (Q(\Delta,t))^n \rangle_c = \int dz M(z) H_n(z) ,
\label{eq:<Q^n>}
\end{align}
with
\begin{align}
H_1(z) =& I_X(z/\Delta) ,
\\
H_2(z) =& I_X(z/\Delta) - I_X(z/\Delta)^2 ,
\\
H_3(z) =& I_X(z/\Delta) - 3 I_X(z/\Delta)^2 + 2 I_X(z/\Delta)^3 ,
\\
H_4(z) =& 
I_X(z/\Delta) - 7 I_X(z/\Delta)^2 + 12 I_X(z/\Delta)^3 
- 6 I_X(z/\Delta)^4 .
\end{align}

\subsection{Uniform initial condition}
\label{sec:uniform}

As a special case of the above result, 
it is instructive to consider the solution of the DME 
for a fixed initial condition with a uniform density,
$n(x,0) = M(x) = M$.
With this initial condition, the average density trivially 
takes a constant value for all $t$.
The fluctuations, however, have nontrivial $t$ dependence even 
for this simple case, because they increase
toward the equilibrated value from zero as $t$ becomes larger.

Substituting the initial condition $M(x) = M$ in 
Eq.~(\ref{eq:<Q^n>}) one obtains
\begin{align}
\langle Q(\Delta,t) \rangle_c 
=& M \Delta,
\label{eq:<Q^1>M} \\
\langle (Q(\Delta,t))^2 \rangle_c 
=& M \Delta \left( 1 - F_2(X) \right) ,
\label{eq:<Q^2>M} \\
\langle (Q(\Delta,t))^3 \rangle_c 
=& M \Delta \left( 1 - 3 F_2(X) + 2 F_3(X) \right) ,
\label{eq:<Q^3>M} \\
\langle (Q(\Delta,t))^4 \rangle_c 
=& M \Delta \left( 1 - 7 F_2(X) + 12 F_3(X) - 6 F_4(X) \right) ,
\label{eq:<Q^4>M}
\end{align}
with 
\begin{align}
F_n(X) = \int dz [I_X(z/\Delta)]^n.
\label{eq:F_n}
\end{align}
Some properties of $F_n(X)$ are given in Appendix~\ref{app:I_X}.
Note that the $\Delta$ and $t$ dependences of the cumulants 
are encoded through $X$; as in Eq.~(\ref{eq:X}), 
$X=0$ corresponds to $t=0$ while $X$ goes to infinity for $t\to\infty$
with a nonzero $\Delta$.

Eq.~(\ref{eq:<Q^1>M}) shows that the average 
$\langle Q \rangle = \langle Q \rangle_c = M\Delta$ 
does not have $t$ dependence.
On the other hand, the second and higher order cumulants, 
Eqs.~(\ref{eq:<Q^2>M}) - (\ref{eq:<Q^4>M}), are 
$t$ dependent.
As discussed in Appendix~\ref{app:I_X}, $F_n(X)$ 
for $n\ge2$ are monotonically decreasing functions of $X$ 
with $F_n(0)=1$ and $\lim_{X\to\infty} F_n(X)=0$.
From these properties, one finds that 
Eqs.~(\ref{eq:<Q^2>M}) - (\ref{eq:<Q^4>M}) vanish 
at $t=0$ as are required by the fixed initial condition, 
while they approach the Poissonian value 
$\langle (Q(\Delta,t))^n \rangle_c = M\Delta$ 
in the large $t$ limit.

\begin{figure}
\begin{center}
\includegraphics[width=.49\textwidth]{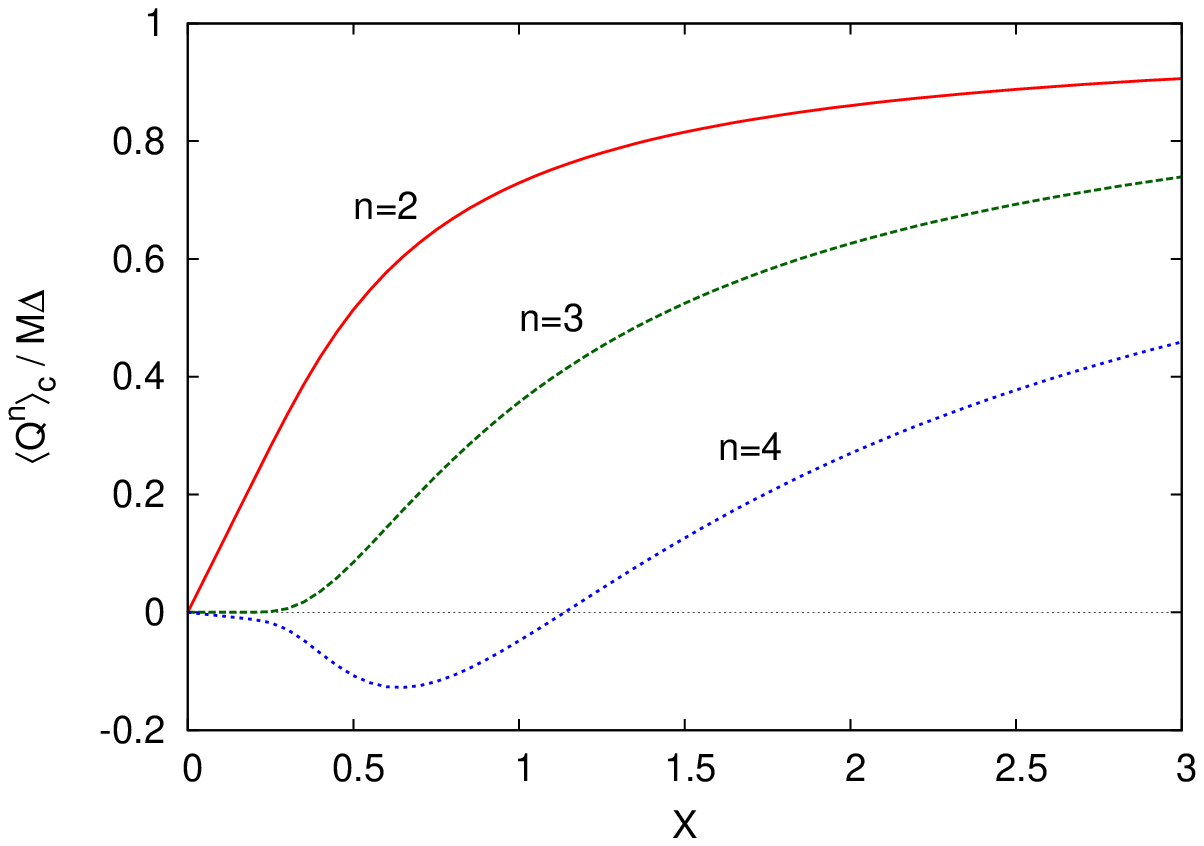}
\includegraphics[width=.49\textwidth]{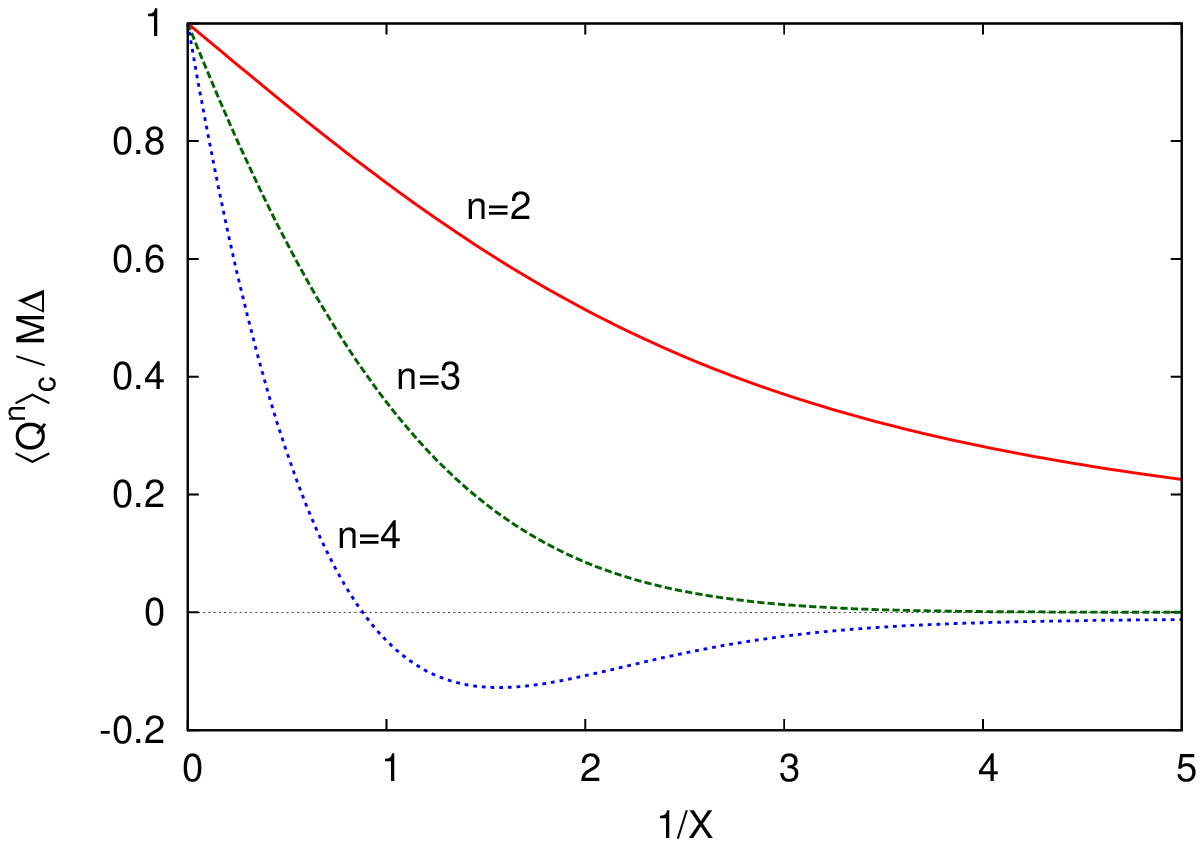}
\caption{
Dependences of the cumulants of $Q(\Delta,\tau)$ on 
$X=d(t)/\Delta$ (left) and $1/X$ (right).
}
\label{fig:Qfixed}
\end{center}
\end{figure}

In Fig.~\ref{fig:Qfixed}, we show the second, third and 
fourth order cumulants normalized by their equilibrated value, 
$M\Delta$, as functions of $X$ and $1/X$.
Because $X$ is a monotonically increasing function of $t$
with fixed $\Delta$,
the left panel can be seen as the time evolution of 
$\langle (Q(\Delta,t))^n \rangle_c$.
The panel shows that the second and third order cumulants 
are monotonically increasing functions.
The fourth order one, on the other hand, has a 
non monotonic time evolution; it first becomes negative,
and then increases toward the equilibrated value.
It is worth emphasizing that the fourth order cumulant can 
become negative owing to the non-equilibrium effect.
One also finds from the figure that the approach of the 
cumulant to the equilibrated value is slower for higher order.
An intuitive interpretation of this tendency is discussed in 
Appendix~\ref{app:chemical}.

In the right panel of Fig.~\ref{fig:Qfixed}, the same 
quantities are shown as functions of $1/X=\Delta/d(t)$.
With fixed $t$, this plot can be regarded as the $\Delta$ 
dependence of $\langle (Q(\Delta,t))^n \rangle_c/M\Delta$.
The figure shows that the equilibration of 
$\langle (Q(\Delta,t))^n \rangle_c$ is realized in the 
$\Delta\to0$ limit, while $\langle (Q(\Delta,t))^n \rangle_c$ 
approaches the initial value as $\Delta$ becomes larger.
In other words, the approach to the equilibrated value 
is faster for smaller $\Delta$.
We note that such a behavior of second-order cumulant of 
net-electric charge fluctuation is 
observed by the ALICE collaboration \cite{ALICE}, 
as will be discussed in more detail in Sec.~\ref{sec:2nd}.

\subsection{General solution}
\label{sec:general}

So far, we have considered the DME Eq.~(\ref{eq:DME}) 
for the fixed initial conditions, 
$n(x,0)=M(x)$ without fluctuations.
Now, we extend these results to general initial 
conditions having fluctuations.
This procedure is nicely carried out by a superposition
of the previous results and using the formula of cumulants 
given in appendix~\ref{app:superposition}.

Let us first denote the probability distribution 
in the initial condition at $t=0$ as 
$P[n(x),0] = F[n(x)]$.
Because Eq.~(\ref{eq:DME}) is a linear equation, 
probability distribution $P[n(x),t]$ with the initial 
distribution $P[n(x),0] = F[n(x)]$ is then given by 
\begin{align}
P[n(x),t] = \sum_{\{M(x)\}} F[M(x)] P_{M(x)} [ n(x),t ],
\label{eq:P[n(x)]}
\end{align}
where $P_{M(x)} [ n(x),t ]$ is the probability distribution 
function at time $t$ with the fixed initial condition with 
$n(z,0)=M(x)$, and the sum runs over the function space of $M(x)$.

Eq.~(\ref{eq:P[n(x)]}) shows that 
$P[n(x),t]$ is given by the superposition of 
probabilities $P_{M(x)} [ n(x),t ]$ with an weight $F[M(x)]$.
In addition, the cumulants of $P_{M(x)} [ n(x),t ]$, 
given in Eq.~(\ref{eq:<Q^n>}), are linear with respect to $M(x)$.
Therefore, the cumulants of Eq.~(\ref{eq:P[n(x)]}) is 
obtained by substituting Eq.~(\ref{eq:<Q^n>}) 
to Eqs.~(\ref{eq:K1NFM}) - (\ref{eq:K4NFM}),
which results in 
\begin{align}
\langle Q(\Delta,t) \rangle_c 
=& \int dz \langle M(z) \rangle_{c,0} H_1(z) ,
\label{eq:<Q^1>0}
\\
\langle (Q(\Delta,t))^2 \rangle_c 
=& \int dz_1 dz_2 
\langle M(z_1) M(z_2) \rangle_{c,0} H_1(z_1) H_1(z_2) 
\nonumber \\
& + \int dz \langle M(z) \rangle_{c,0} 
H_2(z) ,
\label{eq:<Q^2>0}
\\
\langle (Q(\Delta,t))^3 \rangle_c 
=& \int dz_1 dz_2 dz_3
\langle M(z_1) M(z_2) M(z_3) \rangle_{c,0} 
H_1(z_1) H_1(z_2) H_1(z_3) 
\nonumber \\
& + 3 \int dz_1 dz_2 
\langle M(z_1) M(z_2) \rangle_{c,0} 
H_1(z_1) H_2(z_2)
\nonumber \\
& + \int dz \langle M(z) \rangle_{c,0} H_3(z) ,
\label{eq:<Q^3>0}
\\
\langle (Q(\Delta,t))^4 \rangle_c 
=& \int dz_1 dz_2 dz_3 dz_4
\langle M(z_1) M(z_2) M(z_3) M(z_4) \rangle_{c,0} 
\prod_{i=1}^4 H_1(z_i) 
\nonumber \\
& + 6 \int dz_1 dz_2 dz_3
\langle M(z_1) M(z_2) M(z_3) \rangle_{c,0} 
H_1(z_1) H_1(z_2) H_2(z_3)
\nonumber \\
& + \int dz_1 dz_2 
\langle M(z_1) M(z_2) \rangle_{c,0} 
\left\{ 3 H_2(z_1) H_2(z_2) + 4 H_1(z_1) H_3(z_2) \right\}
\nonumber \\
& + \int dz \langle M(z) \rangle_{c,0} H_4(z) .
\label{eq:<Q^4>0}
\end{align}
where $\langle M(z_1) M(z_2) \cdots \rangle_{c,0}$ represents 
the cumulants of the initial distribution $F[M(z)]$.
Note that the above results up to second order are 
consistent with the solution of the stochastic 
diffusion equation.

\subsection{Initial condition satisfying uniformity and locality}
\label{locality}

Using Eqs.~(\ref{eq:<Q^1>0}) - (\ref{eq:<Q^4>0}) one can 
obtain cumulants of $Q(\Delta,t)$ with an arbitrary initial condition
having fluctuations.
We next constrain the initial condition
so that the result becomes physically apparent.

In an equilibrated medium and in a macroscopic scale, 
density fluctuations are local, i.e.
the cumulants of the particle density is given by 
\begin{align}
\langle n(x_1) n(x_2) \cdots n(x_l) \rangle_c
=& [n^l]_c \delta( x_1 - x_2 ) \cdots \delta( x_1 - x_l ),
\label{eq:<n>delta}
\end{align}
where the coefficients in Eq.~(\ref{eq:<n>delta}) defined by 
\begin{align}
[n^l]_c= \langle Q^n \rangle_c/\Delta,
\label{eq:[]}
\end{align}
represent the generalized susceptibility.
This is because the correlation length of the density 
would be at most the same order as the microscopic scales 
such as the mean free path without long range correlations.
The correlation between macroscopically separated points,
therefore, must vanishes.
It is notable that Eq.~(\ref{eq:<n>delta}) is also 
satisfied in the classical free gas in grand canonical 
ensemble.
When, and only when, the cumulants satisfy the condition
Eq.~(\ref{eq:<n>delta}), $\langle Q^n \rangle_c$ are extensive 
variables and proportional to $\Delta$,
\begin{align}
\langle (Q(\Delta,t))^n \rangle_c = \int_0^{\Delta} dx_1 \cdots dx_n
\langle n(x_1) n(x_2) \cdots n(x_l) \rangle_{c,0} 
= \Delta [n^l]_c .
\label{eq:Q[]}
\end{align}
It is reasonable from Eq.~(\ref{eq:Q[]}) to refer to 
$[n^l]_c$ as the generalized susceptibility.
Note that the translational symmetry of the system is 
assumed in Eq.~(\ref{eq:<n>delta});
extension of Eq.~(\ref{eq:<n>delta}) to the finite volume 
system is addressed in Ref.~\cite{SAK}.

For initial conditions satisfying Eq.~(\ref{eq:<n>delta}), 
the cumulants Eqs.~(\ref{eq:<Q^1>0})
- (\ref{eq:<Q^4>0}) are reduced to be
\begin{align}
\langle Q(\Delta,t) \rangle_c 
=& \Delta [M]_c,
\label{eq:<Q^1>L}
\\
\langle (Q(\Delta,t))^2 \rangle_c 
=& 
\Delta [M]_c \left\{ 1 - F_2(X) \right\}
+ \Delta [M^2]_c F_2(X) ,
\label{eq:<Q^2>L}
\\
\langle (Q(\Delta,t))^3 \rangle_c 
=& \Delta [M]_c \left\{ 1 - 3 F_2(X) + 2 F_3(X) \right\}
+ 3 \Delta [M^2]_c \left\{ F_2(X) - F_3(X) \right\}
\nonumber \\
& + \Delta [M^3]_c F_3(X) ,
\label{<eq:Q^3>L}
\\
\langle (Q(\Delta,t))^4 \rangle_c 
=& \Delta [M]_c \left\{ 1 - 7 F_2(X) + 12 F_3(X) - 6 F_4(X) \right\}
\nonumber \\
& + \Delta [M^2]_c \left\{ 7 F_2(X) - 18 F_3(X) + 11 F_4(X) \right\}
\nonumber \\
& + 6 \Delta [M^3]_c \left\{ F_3(X) - F_4(X) \right\}
+ \Delta [M^4]_c F_4(X) ,
\label{<eq:Q^4>L}
\end{align}
where $[M^n]_c$ are the generalized susceptibilities 
for the initial condition satisfying
\begin{align}
[M^n]_c = \frac1\Delta \bigg\langle \bigg( 
\int_0^\Delta dz M(z) \bigg)^n \bigg\rangle .
\end{align}

\section{Extension to multi-particle systems}
\label{sec:multi}

\subsection{Cumulants of net numbers}
\label{sec:net}

Next, let us extend our model to a system composed of 
two species of Brownian particles.
We then consider the cumulants of the difference of 
the particle numbers.
This extension is needed for the description the 
cumulants of the net-baryon and net-electric charge numbers, 
which are defined by the difference of particle numbers.

We denote the densities of two particle species 
as $n_1(x)$ and $n_2(x)$, and assume that their time 
evolutions are given by Eq.~(\ref{eq:DME}).
We then consider the cumulants of the 
difference of the particle numbers in a range of width $\Delta$,
\begin{align}
Q_{\rm net}(\Delta,t) =& Q_1(\Delta,t) - Q_2(\Delta,t) ,
\\
Q_i(\Delta,t) =& \int_{-\Delta/2}^{\Delta/2} dx n_i(x,t) .
\end{align}
One easily finds that the distribution 
of $Q_{\rm net}$ approaches the Skellam one 
in the $t\to\infty$ limit, because in this limit 
the distributions of $Q_1$ and $Q_2$ are given by Poissonian 
and uncorrelated.

For clarity now we recover the discretized 
notation for the particle density for the moment.
We write the probability distribution function
that a cell labeled by $m$ have $n^{(1)}_m$ and $n^{(2)}_m$
particles at time $t$ as $P(\bm{n}^{(1)},\bm{n}^{(2)},t)$.
Because $n^{(1)}_m$ and $n^{(2)}_m$ separately obey
Eq.~(\ref{eq:DME}), if these two densities have no 
correlation in the initial condition, 
$n^{(1)}_m$ and $n^{(2)}_m$ are independent for any $t$.
In this case, the probability can be factorized as 
\begin{align}
P(\bm{n}^{(1)},\bm{n}^{(2)},t)
= P_1(\bm{n}^{(1)},t) P_2(\bm{n}^{(2)},t) .
\label{eq:P=PP}
\end{align}
When $P(\bm{n}^{(1)},\bm{n}^{(2)},t)$ is factorized in this way,
the cumulant generating function 
$K(\bm{\theta}_1,\bm{\theta}_2,t)$ defined by 
\begin{align}
K(\bm{\theta}^{(1)},\bm{\theta}^{(2)},t)
= \sum_{\bm{\theta}^{(1)},\bm{\theta}^{(2)}}
\prod_m e^{ \theta^{(1)}_m n^{(1)}_m + \theta^{(2)}_m n^{(2)}_m }
P(\bm{n}^{(1)},\bm{n}^{(2)},t) ,
\label{eq:K12}
\end{align}
is also decomposed as 
\begin{align}
K(\bm{\theta}_1,\bm{\theta}_2,t)
= K_1(\bm{\theta}_1,t) + K_2(\bm{\theta}_2,t),
\label{eq:K=K1+K2}
\end{align}
where $K(\bm{\theta},t)$ is the cumulant generating 
function for the distribution function of each particle.

Using the generating function Eq.~(\ref{eq:K12}), 
cumulants of the difference of particle numbers 
$\bar{n}_m = n^{(1)}_m - n^{(2)}_m$ 
are given by 
\begin{align}
\langle \bar{n}_{m_1} \bar{n}_{m_2} \cdots \bar{n}_{m_l} \rangle_c
= \frac{ \partial^l K }{ \partial \bar{\theta}_{m_1}
\partial \bar{\theta}_{m_2} \cdots \partial \bar{\theta}_{m_l} },
\label{eq:n=K12}
\end{align}
where $\partial/\partial \bar{\theta}
= \partial/\partial \bar{\theta}_1 
- \partial/\partial \bar{\theta}_2$.
By combining Eqs.~(\ref{eq:n=K12}) and (\ref{eq:K=K1+K2}),
one finds 
\begin{align}
\langle \bar{n}_{m_1} \bar{n}_{m_2} \cdots \bar{n}_{m_l} \rangle_c
=& \langle n^{(1)}_{m_1} n^{(1)}_{m_2} \cdots n^{(1)}_{m_4} \rangle_c
+ (-1)^l \langle n^{(2)}_{m_1} n^{(2)}_{m_2} \cdots n^{(2)}_{m_4} \rangle_c .
\label{eq:n_12}
\end{align}
In the continuum notation and after taking the integral 
over the range $\Delta$, one obtains
\begin{align}
\langle (Q_{\rm net}(\Delta,t))^n \rangle_c
=& \langle (Q_1(\Delta,t))^n \rangle_c 
+ (-1)^n \langle (Q_2(\Delta,t))^n \rangle_c ,
\label{eq:Q(-)n}
\end{align}
for initial conditions without correlation between two particle
species.

When the fluctuations of $n_1(x)$ and $n_2(x)$ are correlated 
in the initial condition, the probability distribution is 
no longer factorized as in Eq.~(\ref{eq:P=PP}).
In this case, we denote the probability distribution 
as a superposition of the solutions of fixed initial
conditions,
\begin{align}
P[n_1(x),n_2(x),t]
&= \sum_{\{M_1(x),M_2(x)\}} F[M_1(z),M_2(z)] 
P_{M_1(z),M_2(z)}[n_1(x),n_2(x),t] \nonumber \\
&= \sum_{\{M_1(x),M_2(x)\}} F[M_1(z),M_2(z)] 
P_{M_1(z)}[n_1(x),t] 
P_{M_2(z)}[n_2(x),t] ,
\label{eq:P12}
\end{align}
where $P_{M_1(z),M_2(z)}[n_1(x),n_2(x),t]$ are the solutions 
for the fixed initial condition with 
$n_1(x,0)=M_1(x)$ and $n_2(x,0)=M_2(x)$.
In the second line of Eq.~(\ref{eq:P12}) we used the 
fact that $n_1(x)$ and $n_2(x)$ are uncorrelated 
with the fixed initial condition.
It is also concluded from this uncorrelated nature that 
the cumulants of $P_{M_1(z),M_2(z)}[n_1(x),n_2(x),t]$ 
satisfy Eq.~(\ref{eq:Q(-)n}).
Using Eq.~(\ref{eq:Q(-)n}) and the relation for superposition 
of probabilities in appendix~\ref{app:superposition}, 
the cumulants of $Q_{\rm net}(\Delta,t)$ are given by,
\begin{align}
\langle Q_{\rm net}(\Delta,t) \rangle_c 
=& \int dz \langle M_{\rm net}(z) \rangle_{c,0} H_1(z) ,
\label{eq:<Q_1>net}
\\
\langle (Q_{\rm net}(\Delta,t))^2 \rangle_c 
=& \int dz_1 dz_2 
\langle M_{\rm net}(z_1) M_{\rm net}(z_2) \rangle_{c,0} H_1(z_1) H_1(z_2) 
\nonumber \\
& + \int dz \langle M_{\rm tot}(z) \rangle_{c,0} 
H_2(z) ,
\label{eq:<Q_2>net}
\\
\langle (Q_{\rm net}(\Delta,t))^3 \rangle_c 
=& \int dz_1 dz_2 dz_3
\langle M_{\rm net}(z_1) M_{\rm net}(z_2) M_{\rm net}(z_3) \rangle_{c,0} 
H_1(z_1) H_1(z_2) H_1(z_3) 
\nonumber \\
& + 3 \int dz_1 dz_2 
\langle M_{\rm net}(z_1) M_{\rm tot}(z_2) \rangle_{c,0} 
H_1(z_1) H_2(z_2)
\nonumber \\
& + \int dz \langle M_{\rm net}(z) \rangle_{c,0} H_3(z) ,
\label{eq:<Q_3>net}
\\
\langle (Q_{\rm net}(\Delta,t))^4 \rangle_c 
=& \int dz_1 dz_2 dz_3 dz_4
\langle M_{\rm net}(z_1) M_{\rm net}(z_2) M_{\rm net}(z_3) 
M_{\rm net}(z_4) \rangle_{c,0} 
\prod_{i=1}^4 H_1(z_i) 
\nonumber \\
& + 6 \int dz_1 dz_2 dz_3
\langle M_{\rm net}(z_1) M_{\rm net}(z_2) M_{\rm tot}(z_3) \rangle_{c,0} 
H_1(z_1) H_1(z_2) H_2(z_3)
\nonumber \\
& + 3 \int dz_1 dz_2 
\langle M_{\rm tot}(z_1) M_{\rm tot}(z_2) \rangle_{c,0} 
H_2(z_1) H_2(z_2)
\nonumber \\
& + 4 \int dz_1 dz_2 
\langle M_{\rm net}(z_1) M_{\rm net}(z_2) \rangle_{c,0} 
H_1(z_1) H_3(z_2)
\nonumber \\
& + \int dz \langle M_{\rm tot}(z) \rangle_{c,0} H_4(z) ,
\label{eq:<Q_4>net}
\end{align}
with $M_{\rm net}(z) = M_1(z) - M_2(z)$ and 
$M_{\rm tot}(z) = M_1(z) + M_2(z)$.

When the initial condition satisfies the locality condition
Eq.~(\ref{eq:<n>delta}), cumulants of $Q_{\rm net}(\Delta,t)$ are
\begin{align}
\langle Q_{\rm net}(\Delta,t) \rangle_c 
=& \Delta [M_{\rm net}]_c,
\label{eq:<Q^1>Lnet}
\\
\langle (Q_{\rm net}(\Delta,t))^2 \rangle_c 
=& 
\Delta [M_{\rm tot}]_c \left\{ 1 - F_2(X) \right\}
+ \Delta [M_{\rm net}^2]_c F_2(X) ,
\label{eq:<Q^2>Lnet}
\\
\langle (Q_{\rm net}(\Delta,t))^3 \rangle_c 
=& \Delta [M_{\rm net}]_c \left\{ 1 - 3 F_2(X) + 2 F_3(X) \right\}
+ 3 \Delta [M_{\rm net} M_{\rm tot}]_c 
\left\{ F_2(X) - F_3(X) \right\}
\nonumber \\
& + \Delta [M_{\rm net}^3]_c F_3(X) ,
\label{eq:<Q^3>Lnet}
\\
\langle (Q_{\rm net}(\Delta,t))^4 \rangle_c 
=& \Delta [M_{\rm tot}]_c \left\{ 1 - 7 F_2(X) + 12 F_3(X) - 6 F_4(X) \right\}
\nonumber \\
& + 3 \Delta [(M_{\rm tot})^2]_c 
\left\{ F_2(X) - 2 F_3(X) + F_4(X) \right\}
\nonumber \\
& + 4 \Delta [M_{\rm net}^2]_c 
\left\{ F_2(X) - 3 F_3(X) + 2 F_4(X) \right\}
\nonumber \\
& + 6 \Delta [M_{\rm net}^2 M^{\rm tot}]_c 
\left\{ F_3(X) - F_4(X) \right\}
+ \Delta [M_{\rm net}^4]_c F_4(X) .
\label{eq:<Q^4>Lnet}
\end{align}
Here, $[(M_{\rm net})^i (M_{\rm tot})^j]_c$ are generalized 
susceptibilities defined similarly to Eq.~(\ref{eq:[]}).

The cumulants in equilibrium in this system is obtained by 
taking the large $t$ limit in Eqs.~(\ref{eq:<Q^1>Lnet}) - 
(\ref{eq:<Q^4>Lnet}).
By taking $X\to\infty$, $F_n(X)$ for $n\ge2$ vanish and 
the cumulants of $Q_{\rm net}$ become the Skellam ones with
\begin{align}
\langle (Q_{\rm net})^{2i+1} \rangle_c = \Delta [M_{\rm net}]_c , \quad
\langle (Q_{\rm net})^{2i} \rangle_c = \Delta [M_{\rm tot}]_c ,
\label{eq:QSkellam}
\end{align}
with $i$ being an integer.

\subsection{Generalization to multi-particle systems}

The above result would be suitable to describe the 
time evolution of net-baryon number cumulants in the hadronic 
medium in heavy ion collisions, because the net-baryon number, 
$N_{\rm B}^{\rm net}$,
is given by the difference of baryon and anti-baryon numbers, 
$N_{\rm B}$ and $N_{\bar{\rm B}}$, respectively, as 
\begin{align}
N_{\rm B}^{\rm net} = N_{\rm B} - N_{\bar{\rm B}}.
\end{align}

When one considers the cumulants of net-electric charge, 
there are hadrons having $\pm2$ charge in the unit of elementary 
charge in addition to those having $\pm1$.
The net-electric charge is thus given by 
\begin{align}
N_{\rm Q}^{\rm net} = 2 N_{\rm Q,2} + N_{\rm Q,1}
- N_{\rm Q,-1} - 2 N_{\rm Q,-2},
\label{eq:N_Q}
\end{align}
where $N_{{\rm Q},n}$ is the number of hadrons 
with $n$ electric charge.
Owing to the existence of nonzero $N_{\rm Q,\pm2}$,
then, even in the HRG model in which all $N_{{\rm Q},n}$ 
are given by the Poisson distribution and uncorrelated,
the distribution of $N_{\rm Q}^{\rm net}$ deviates from 
the Skellam one.

To describe this non-Skellam property, 
one may extend the DME to include four particle species.
By constructing the net-electric charge as in 
Eq.~(\ref{eq:N_Q}), 
the distribution of the net-electric charge number 
in equilibrium has a deviation from the Skellam one.
The effects of the charge $2$ hadrons, however, are not large 
in the HRG model because of their small abundance \cite{HRG}.
We thus do not consider their effects in this paper 
and left the incorporation of these effects for future study.

\section{Rapidity window dependence of cumulants in heavy ion collisions}
\label{sec:Dy}

Next, using the result obtained in the previous section
we consider the time evolution of the cumulants of conserved 
charges and their $\Delta\eta$ dependence observed 
in relativistic heavy ion collisions.
As for conserved charges, we consider the net-electric charge and 
net-baryon number, which are both observable in these experiments.
Note that the measurement of the net-baryon number
cumulants is possible \cite{KA1,KA2}, although the detectors
cannot observe neutral baryons.

In QCD at sufficiently low temperature and low baryon 
chemical potential, the thermodynamic quantities including 
cumulants of conserved charges are well described by 
the HRG model.
In this model, the distributions of net-baryon number and 
net-electric charge are given by the Skellam distribution
to a good approximation.
In dense and/or hot medium, the cumulants of conserved charges 
are expected to have deviations from the HRG values reflecting 
the change of the medium property including phase transitions 
to the deconfined medium.
For example, 
fluctuations of net-baryon number and net-electric
charge normalized by entropy density or net-baryon number 
are suppressed in the deconfined phase reflecting 
the decrease of the charges carried by quasi-particle excitations 
\cite{Asakawa:2000wh,Jeon:2000wg,Ejiri:2005wq}.
Near the QCD critical point, the cumulants are expected 
to show characteristic suppressions and enhancements 
reflecting the critical phenomena 
\cite{Stephanov:2008qz,Asakawa:2009aj,Stephanov:2011pb}.
The goal of the experimental study of the cumulants of 
conserved charges is to find and confirm these signals
\cite{Koch:2008ia}.

These signals, which would be developed in the 
deconfined medium or near the QCD critical point, however, 
are not directly observable.
Before they are observed by the detectors, the hot medium 
undergoes a time evolution in the hadronic stage.
The fluctuations are modified in the hadronic medium, 
presumably toward the one of the equilibrated hadronic medium.
When one interprets the experimental results on fluctuations,
one must keep this effect in mind; direct comparisons of 
the experimental results on cumulants with a thermal value 
at some early stage in heavy ion collisions would lead to 
a wrong conclusion.
The purpose of this section is to describe the diffusive
process of fluctuations in the hadronic stage and to estimate 
their effects on experimental results using the 
solution of the DME obtained in previous sections.

In relativistic heavy ion collisions with sufficiently
large collision energy per nucleon, $\sqrt{s_{\rm NN}}$, 
the hot medium created by the collisions has an approximate 
boost invariance around mid-rapidity.
Useful coordinates to describe such a system are the 
coordinate-space rapidity $y=(1/2)\log(t+x)(t-x)$ and 
proper time $\tau=\sqrt{t^2-x^2}$, where $x$ denotes 
the longitudinal direction.
In this coordinate, the diffusion equation 
in the Cartesian coordinate, Eq.~(\ref{eq:diffusion}), 
is rewritten as 
\begin{align}
\frac{\partial}{\partial\tau} n(\eta,\tau)
= D_y(\tau) \frac{\partial^2}{\partial y^2} n(y,\tau),
\label{eq:diffusion_eta}
\end{align}
where $n(y,\tau)$ is the density per unit rapidity, 
and the diffusion constant in this coordinate, $D_y(\tau)$,
is related to the Cartesian one $D$ as
\begin{align}
D_y(\tau) = D \tau^{-2}.
\label{eq:D:eta-tau}
\end{align}
Because of Eq.~(\ref{eq:diffusion_eta}),
the diffusion constant $D$ has to be replaced by $D_y$
when one applies the result in the previous sections 
to the description of the diffusive process along 
the rapidity direction.

In experiments, the fluctuations of particle numbers 
in a pseudo-rapidity window $\Delta\eta$ are observed after 
integrating out the transverse direction \cite{STAR,ALICE}.
The pseudo-rapidity window can be varied within the 
coverage of the detector.
In the following, we regard the pseudo-rapidity window 
as the one of the coordinate space rapidity $\eta$ \cite{OAKS}.
The $\Delta\eta$ dependence of the cumulants 
can then be compared with the $\Delta$ dependence 
obtained in previous sections.
The cumulants obtained in the previous section,
Eqs.~(\ref{eq:<Q^1>Lnet}) - (\ref{eq:<Q^4>Lnet}), depends 
on the initial condition and the diffusion distance $d(t)$.
By fitting the experimental data by these results,
one can constrain the initial condition and the value of $d(t)$,
namely information on the thermodynamics and transport property
of the hot medium.

In heavy ion collisions, the conserved charges in the total 
system is fixed and does not fluctuate.
Because the hot medium created in heavy ion collisions is 
a finite volume system, the measurement of fluctuation 
in a sub-volume is affected by this effect.
For sufficiently large $\sqrt{s_{\rm NN}}$ and moderate range 
of $\Delta\eta$, however, such an effect is suppressed 
because of the finite duration of the diffusive process 
in the hadronic medium \cite{SAK}.

\subsection{Initial condition}
\label{sec:initial}

In this study, we set the medium at chemical freezeout
time, $\tau=\tau_0$, as the initial condition, and apply 
the DME to describe the diffusive process 
in the hadronic medium until kinetic freezeout.
For sufficiently large $\sqrt{s_{\rm NN}}$, hadronization 
and chemical freezeout are expected to take place at almost 
the same time.
Because the local density profile of conserved charges in 
each event is almost frozen within the short duration 
because of the charge conservation, 
if the fluctuations are well equilibrated in the early stage, 
the fluctuations of conserved charges at $\tau=\tau_0$ will 
strongly reflect the thermal property in the deconfined medium.
The fluctuations of non-conserving quantities, on the 
other hand, are not constrained by the conservation law, 
and thus would be sensitive to the hadronization mechanism
rather than the primordial thermodynamics.

In this study, we assume that the locality condition 
Eq.~(\ref{eq:<n>delta}) is satisfied at $\tau=\tau_0$.
Note that this assumption is satisfied if the medium 
is fully equilibrated at this time.
With this assumption, the initial condition is 
specified by the generalized susceptibilities.
To describe the cumulants up to fourth order, we need 
six susceptibilities in the initial condition
\begin{align}
[M_{\rm net}^2]_c, \quad 
[M_{\rm net}^3]_c, \quad 
[M_{\rm net}^4]_c, \quad
[M_{\rm net}M_{\rm tot}]_c, \quad 
[M_{\rm net}^2 M_{\rm tot}]_c, \quad
[(M_{\rm tot})^2]_c.
\end{align}
The first three susceptibilities are theose
of conserved charges.
By normalizing them by their equilibrated values 
given in Eq.~(\ref{eq:QSkellam}),
we introduce the following three parameters 
\begin{align}
D_2 = [M_{\rm net}^2]_c/[M_{\rm tot}]_c , \quad
D_3 = [M_{\rm net}^3]_c/[M_{\rm net}]_c , \quad
D_4 = [M_{\rm net}^4]_c/[M_{\rm tot}]_c, 
\label{eq:D234}
\end{align}
representing the magnitude of the cumulants of 
conserved charges at hadronization.
Other three parameters are not the cumulants of conserved charges.
We parametrize these quantities by 
\begin{align}
a = [M_{\rm net}M_{\rm tot}]_c/[M_{\rm net}]_c , \quad
b = [M_{\rm net}^2 M_{\rm tot}]_c/[M_{\rm tot}]_c , \quad
c = [(M_{\rm tot})^2]_c/[M_{\rm tot}]_c , 
\label{eq:abc}
\end{align}
where we again normalize them by the equilibrated values.
Among these parameters, $D_n$ are the quantities 
which should be compared with theoretical 
studies, such as lattice simulations, on the cumulants of 
conserved charges.
Note that $D_2$ and $c$ are positive definite because they 
are second-order fluctuations, while other parameters can 
take both positive and negative values.
In addition to Eqs.~(\ref{eq:D234}) and (\ref{eq:abc}), 
the $\Delta\eta$ dependence of the cumulants depends on the the 
diffusion length $d(\tau)$.

Because the parameters in Eq.~(\ref{eq:abc}) are not directly 
constrained by the conservation law, their values would be 
sensitive to the hadronization mechanism rather than the 
primordial thermodynamics.
In particular, the parameter $c$ does not include the 
net-particle number and is completely free from the 
conservation law.
Experimental constraint on this parameter would provide
us with novel information on the hadronization mechanism 
\cite{KAO}.

For net-baryon number cumulants and for sufficiently low 
energy collisions at which the creation of the anti-baryons 
are well suppressed, the system can be regarded as the single 
particle one without anti-baryons.
Because $M_{\rm net}$ and $M_{\rm tot}$ become identical 
in this case, one obtains $a=c=D_2$ and $b=D_3$.
The number of parameters are reduced in this case.

\subsection{Normalized cumulants}
\label{sec:normalized}

In the following, we plot the $\Delta\eta$ dependences of 
cumulants by normalizing by their equilibrated values, 
\begin{align}
R_n(X) = \frac{ \langle (Q_{\rm net}(\Delta\eta,\tau))^n \rangle_c }
{ \lim_{\tau\to\infty} \langle (Q_{\rm net}(\Delta\eta,\tau))^n \rangle_c }.
\label{eq:R}
\end{align}
We refer to Eq.~(\ref{eq:R}) as the normalized cumulants
in what follows.
Note that the equilibrated values are the Skellam one 
given in Eq.~(\ref{eq:QSkellam}) in our model.

When one analyzes the normalized cumulants in experiments, 
one may use the equilibrated values in the HRG model for normalization.
An alternative way to deduce the equilibrated value 
is to take the $\Delta\eta\to0$ limit of the experimental results 
on the cumulant.

Usually, the experimental results on the higher order cumulants 
are discussed in terms of their ratios such as $R_4/R_2$ 
and $R_3/R_1$ \cite{STAR}.
When one considers the $\Delta\eta$ dependences, however, 
the normalized cumulants would be better than the ratios,
since one can investigate the individual cumulants directly 
using $R_n$.

\begin{figure}
\begin{center}
\includegraphics[width=.45\textwidth]{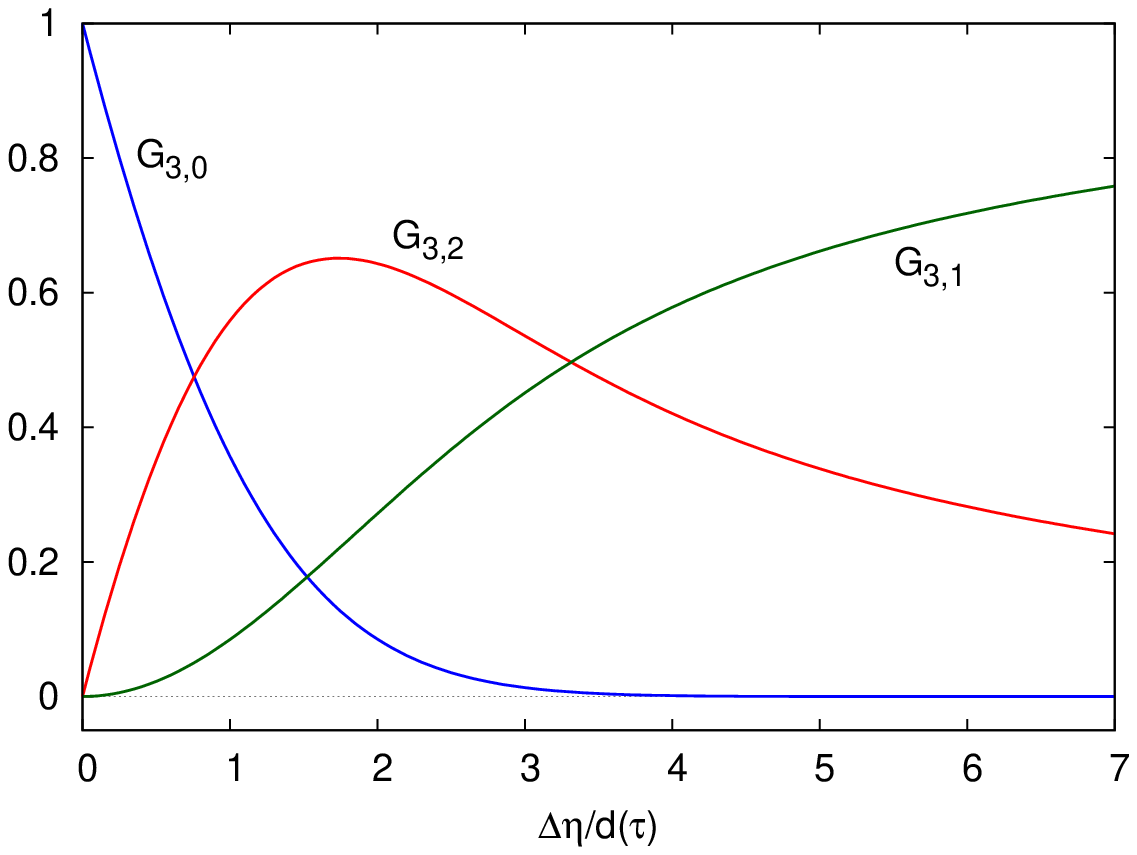}
\includegraphics[width=.45\textwidth]{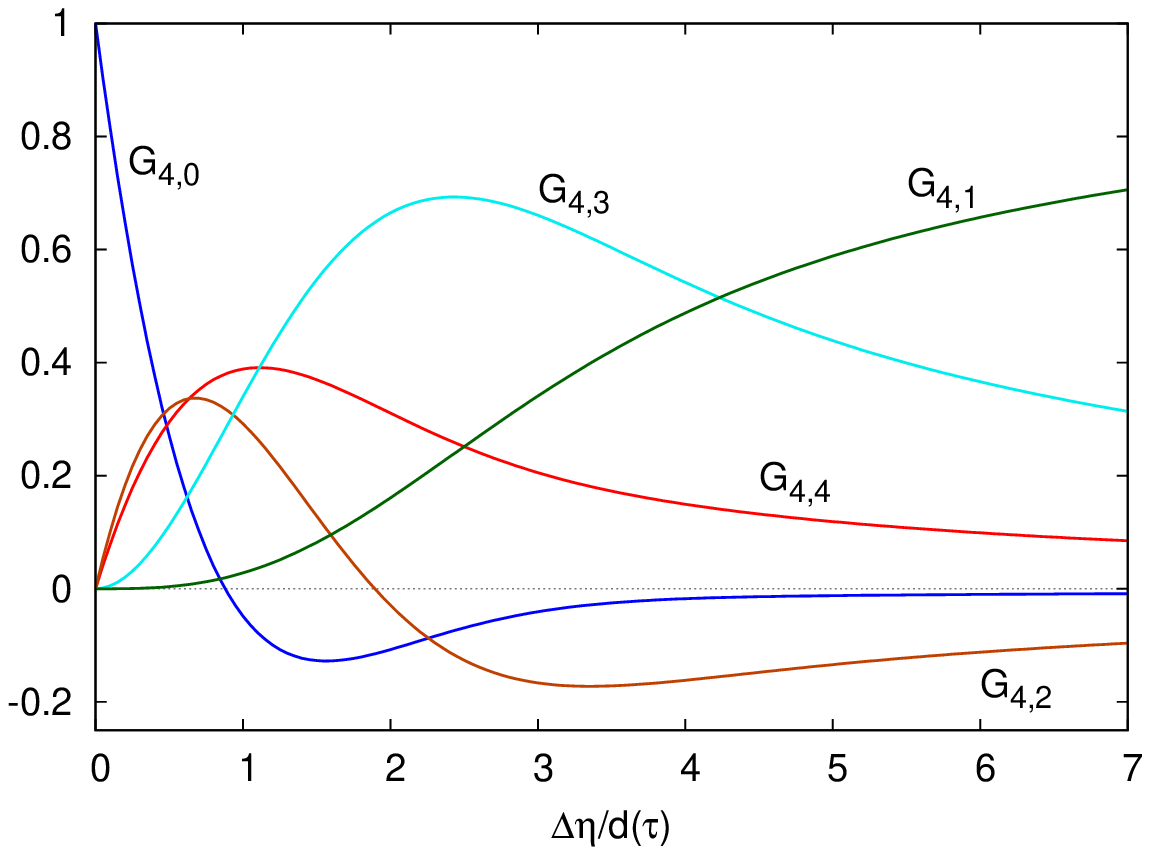}
\caption{
$\Delta\eta/d(\tau)$ dependences of the 
functions defined in Eqs.~(\ref{eq:G3}) - (\ref{eq:G4}).
}
\label{fig:G_elem}
\end{center}
\end{figure}

Using the parameters in Eqs.~(\ref{eq:D234}) and (\ref{eq:abc}),
the solution obtained in the previous section 
Eqs.~(\ref{eq:<Q^1>Lnet}) - (\ref{eq:<Q^4>Lnet}) 
in terms of the normalized cumulants are given by 
\begin{align}
R_2(X) &= G_{2,0}(X) + D_2 G_{2,1}(X),
\label{eq:R2}
\\
R_3(X) &= G_{3,0}(X) + D_3 G_{3,1}(X) + a G_{3,2}(X),
\label{eq:R3}
\\
R_4(X) &= G_{4,0}(X) + D_4 G_{4,1}(X) + D_2 G_{4,2}(X)
+ b G_{4,3}(X) + c G_{4,4}(X),
\label{eq:R4}
\end{align}
where we have introduced the following functions:
\begin{align}
G_{2,0}(X) &= 1 - F_2(X) , \quad 
G_{2,1}(X) = F_2(X), 
\label{eq:G2}
\\
G_{3,0}(X) &= 1 - 3 F_2(X) + 2 F_3(X) , \quad 
G_{3,1}(X) = F_3(X) ,  \\
G_{3,2}(X) &= 3( F_2(X) - F_3(X) ),
\label{eq:G3}
\\
G_{4,0}(X) &= 1 - 7 F_2(X) + 12 F_3(X) - 6 F_4(X) , \quad
G_{4,1}(X) = F_4(X) ,  \\
G_{4,2}(X) &= 4( F_2(X) - 3 F_3(X) + 2 F_4(X) ) ,
\\
G_{4,3}(X) &= 6( F_3(X) - F_4(X) ) , \quad
G_{4,4}(X) = 3( F_2(X) - 2 F_3(X) + F_4(X) ).
\label{eq:G4}
\end{align}
In Fig.~\ref{fig:G_elem}, we show $1/X=\Delta\eta/d(\tau)$ 
dependence of the functions in Eqs.~(\ref{eq:G2}) - (\ref{eq:G4}).
The figure shows that these functions behave 
differently as functions of $\Delta\eta/d(\tau)$.
The structures of these functions are responsible for 
characteristic behaviors of $\Delta\eta$ dependences of the 
cumulants discussed in the next subsections.

\subsection{Second order cumulant and diffusion distance}
\label{sec:2nd}

\begin{figure}
\begin{center}
\includegraphics[width=.45\textwidth]{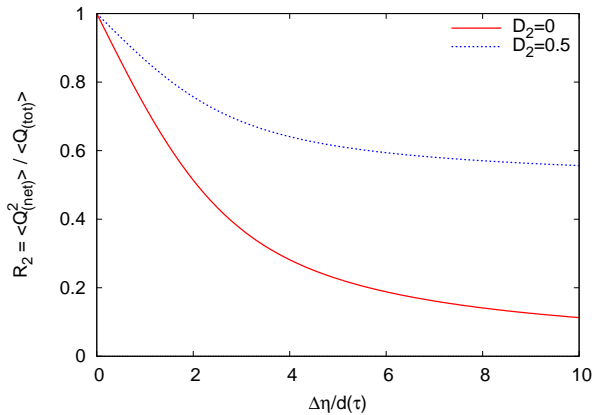}
\caption{
$\Delta\eta/d(\tau)=1/X$ dependence of the normalized 
second order cumulant $R_2$ for $D_2=0$ and $0.5$.
}
\label{fig:R2}
\end{center}
\end{figure}

Now, let us examine $\Delta\eta/d(\tau)=1/X$ dependence of 
the normalized cumulants Eq.~(\ref{eq:R}).
We first consider the the second order one $R_2$.
As in Eq.~(\ref{eq:R2}), $R_2$ depends 
on the initial condition only through $D_2$.
In Fig.~\ref{fig:R2} we show the $\Delta\eta/d(\tau)$ 
dependence of $R_2$ for $D_2=0$ and $0.5$.

Since we plot the $\Delta\eta$ dependences of the cumulants
as functions of $\Delta\eta/d(\tau)$ throughout this section,
it is instructive to give a rough estimate on 
the value of $d(\tau)$ at this point by comparing 
Fig.~\ref{fig:R2} with the existing experimental results.
The second order cumulant of net-electric charge 
has been observed by ALICE collaboration \cite{ALICE}.
In Ref.~\cite{ALICE}, the result is plotted using the 
quantity called the $D$-measure \cite{Jeon:2000wg}, 
which is related to $R_2$ as
\begin{align}
D = 4 R_2,
\end{align}
provided that $Q_{\rm net}$ is the net-electric charge. 
By comparing the result in Fig.~\ref{fig:R2} with 
Fig.~3 in Ref.~\cite{ALICE}, one can constrain the 
values of $D_2$ and $d(\tau)$ \cite{SAK}.
In Ref.~\cite{Asakawa:2000wh},
the value of $D_2$ is estimated as $D_2=0.5$.
Using this value of $D_2$ and Fig.~3 in Ref.~\cite{ALICE}
one can estimate \cite{SAK}
\begin{align}
d(\tau) = 0.3 \sim 0.5.
\end{align}
This result shows that the maximum rapidity window
of the ALICE detector, $\Delta\eta_{\rm max}=1.6$ \cite{ALICE},
corresponds to 
\begin{align}
\Delta\eta_{\rm max}/d(\tau) = 3.2 \sim 5.3.
\end{align}
The ALICE detector thus can analyze $\Delta\eta/d(\tau)$ 
dependences of cumulants for 
$0<\Delta\eta<\Delta\eta_{\rm max}$.
Note that this estimate on $d(\tau)$ strongly 
depends on the value of $D_2$; for smaller value of 
$D_2$, $d(\tau)$ becomes much larger.

When one applies the net-baryon number to $Q_{\rm net}$,
the value of of $d(\tau)$ should be much smaller 
because the hadrons having baryon number, namely baryons,
are considerably heavier than those having electric charge,
which are dominated by pions in heavy ion collisions.
This suggests that wider range of $\Delta\eta/d(\tau)$ 
can be analyzed for net-baryon number cumulants with 
a fixed $\Delta\eta$ coverage of an experimental detector.
The use of the net-baryon number cumulants is advantageous 
compared with net-electric charge in this sense.
The simultaneous measurements of both the net-baryon
number and net-electric charge cumulants in a same experiment 
would provide us interesting results, such as the difference
of the diffusion constants of net-baryon number and net-electric
charge.

\subsection{$\Delta\eta$ dependence of higher order cumulants}
\label{sec:higher}

\subsubsection{Initial condition with small fluctuations}

\begin{figure}
\begin{center}
\includegraphics[width=.45\textwidth]{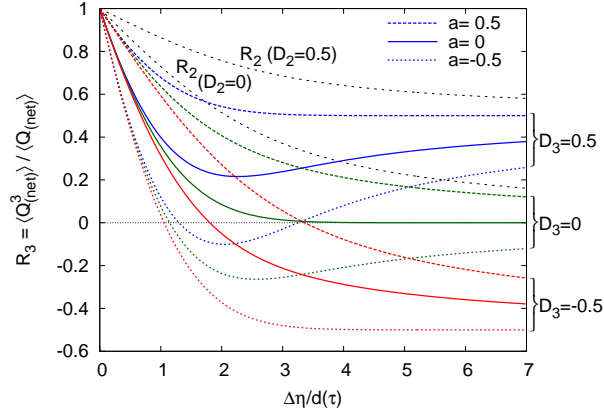}
\caption{
Normalized third-order cumulants $R_3$ as a function of 
$1/X=\Delta\eta/d(\tau)$ for $D_3=-0.5$, $0$, $0.5$ and 
several values of $a$.
}
\label{fig:R3}
\end{center}
\end{figure}

\begin{figure}
\begin{center}
\includegraphics[width=.45\textwidth]{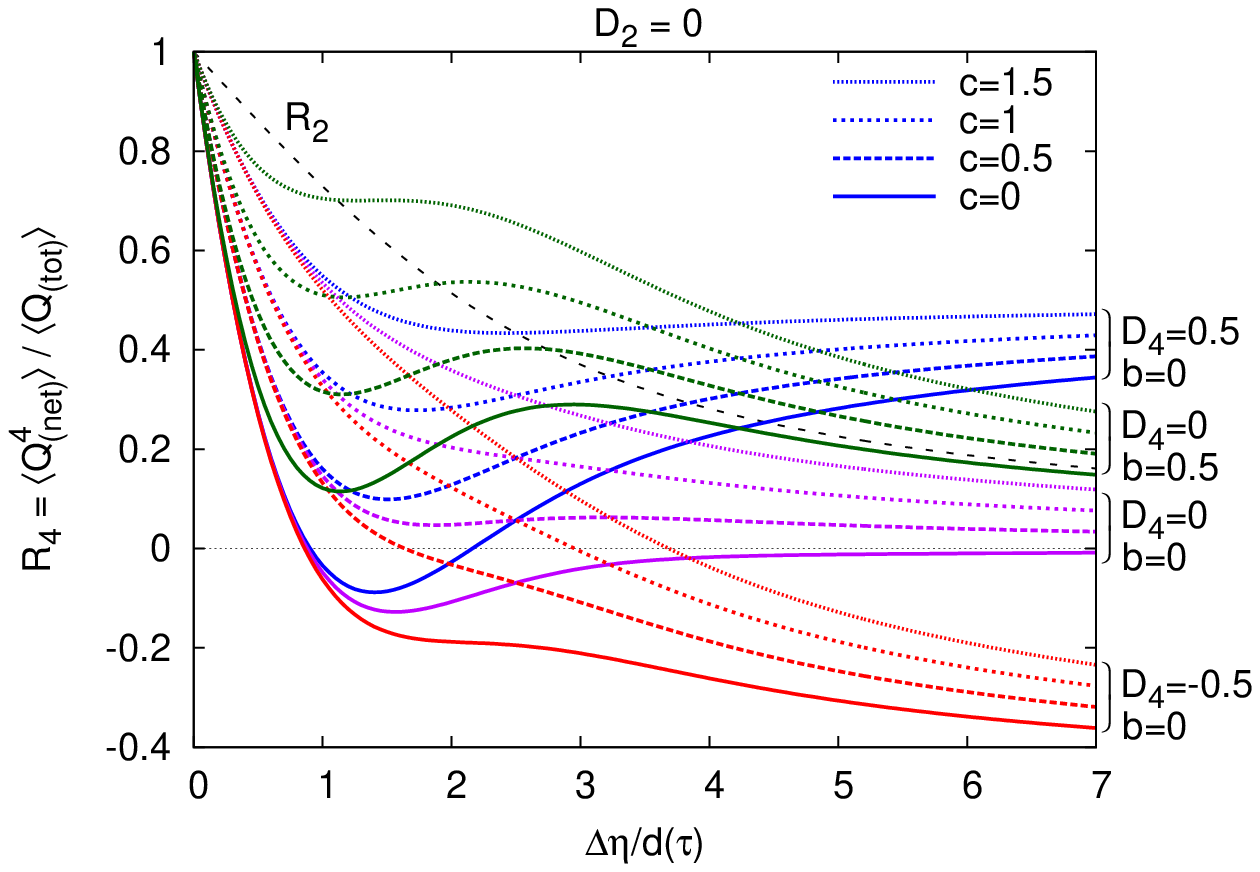}
\includegraphics[width=.45\textwidth]{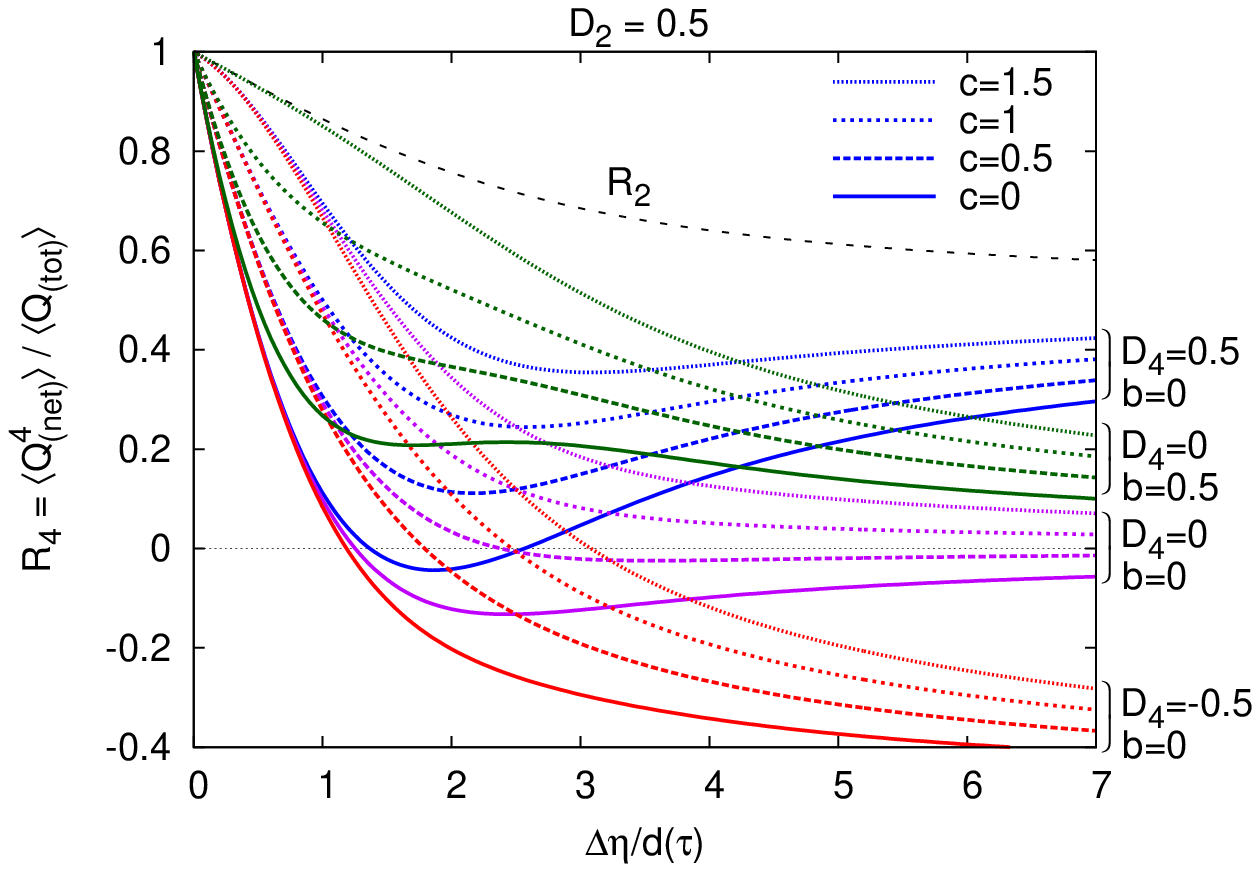}
\caption{
Normalized fourth-order cumulants $R_4$ as a function of 
$\Delta\eta/d(\tau)$ for various initial parameters
$D_4$, $D_2$, $b$ and $c$.
The left (right) panel shows the result for $D_2=0$ ($D_2=0.5$).
}
\label{fig:R4}
\end{center}
\end{figure}

Next, we focus on the $\Delta\eta$ dependences of 
third and fourth order cumulants.
We first consider the initial condition with small 
fluctuations, since the fluctuations of net-baryon and 
net-electric charge numbers are expected to be 
suppressed in the deconfined medium 
\cite{Asakawa:2000wh,Jeon:2000wg,Ejiri:2005wq}.
The higher order cumulants can also be suppressed 
near the QCD critical point reflecting the critical 
phenomena \cite{Asakawa:2009aj,Stephanov:2011pb}.

The normalized cumulant $R_3$ depends on 
$D_3$ and $a$, while $R_4$ depends on 
$D_4$, $D_2$, $b$ and $c$.
In Fig.~\ref{fig:R3}, we show dependence of 
$R_3$ on $\Delta\eta/d(\tau)=1/X$ 
with $D_3=-0.5$, $0$ and $0.5$ for several values of $a$.
In Fig.~\ref{fig:R4}, $\Delta\eta/d(\tau)$ dependence of 
$R_4$ is shown for various values of the initial parameters.
The figures show that the dependences of $R_n$ on 
$\Delta\eta/d(\tau)$ are sensitive to the initial 
parameters in the range $\Delta\eta/d(\tau)\lesssim7$.
For example, one sees from Fig.~\ref{fig:R3} that the behavior 
of $R_3$ is sensitive to $a$ for $\Delta\eta/d(\tau)<1$ 
while the effect of $D_3$ becomes dominant as 
$\Delta\eta/d(\tau)$ becomes larger.
This behavior indicates that the experimental measurement of 
$R_3$ in the range $\Delta\eta/d(\tau)\lesssim5$ makes it 
possible to constrain two parameters $D_3$ and $a$.
It is also notable that $\Delta\eta$ dependence of $R_3$ can 
become non-monotonic for several choices of parameters.
Experimental observation of such a non-monotonic behavior
should be an interesting and unique signal to constrain 
these parameters and underlying physics.
Figure~\ref{fig:R4} also shows that the behavior of 
$R_4$ is sensitive to the initial parameters.
For example, $R_4$ can have two extrema for a choice 
of initial parameter in the range $\Delta\eta/d(\tau)<5$.
Although there are $16$ lines for $R_4$ in each panel of 
Fig.~\ref{fig:R4}, these lines would be experimentally 
distinguished if the experimental result on $R_4$ is obtained 
within $10\%$ precision.
This precision will be achieved 
in the BES-II program at RHIC \cite{BES-II}.
Such experimental analysis can constrain the initial
parameters $D_4$, $b$ and $c$ as well as $d(\tau)$.
We also note that, although $R_4$ depends on $D_2$, 
the value of $D_2$ can be constrained by the measurement 
of $R_2$. 
Similarly, because all normalized cumulants $R_2$, $R_3$ and $R_4$ 
depend on a common $d(\tau)$, this parameter can be constrained 
by the simultaneous use of these three cumulants.
In this way, the complementary use of these cumulants
will lead us to a deeper understanding of the experimental results.
The analyses of both the net-baryon and net-electric charge 
fluctuations in a same set of collision events will also 
give us further interesting information on the difference of 
the diffusion constants for these charges and so forth.

For the analysis of the net-baryon number cumulants with 
small $\sqrt{s_{\rm NN}}$, because the abundance of anti-baryons 
are suppressed the parameters are constrained as 
$a\simeq D_2$, $c\simeq D_2$ and $b\simeq D_3$.
Using these conditions, the analysis of the $\Delta\eta$ 
dependence of normalized cumulants $R_2$, $R_3$ and $R_4$ 
would further constrain the physics of fluctuations.

It is worth emphasizing that 
the results in Figs.~\ref{fig:R3} and \ref{fig:R4} 
show that the value of a cumulant with a fixed 
$\Delta\eta$ can deviate from both the initial and 
equilibrated values significantly.
In particular, the sign of non-Gaussian cumulants can 
become negative even when the initial and equilibrated 
values are positive.
These results show that the cumulants observed in experiments 
with a fixed $\Delta\eta$ should not be compared with the 
theoretical analysis of the cumulants obtained in 
statistical mechanics assuming an equilibration.
Total description of the $\Delta\eta$ dependence
of the cumulants are indispensable for proper understanding
of the experimental results.

\subsubsection{Initial condition with large cumulant}

Next, we show the $\Delta\eta$ dependence of the cumulants
for initial conditions where higher order cumulants of 
conserved charges are consistent with or larger than 
the equilibrated values.

\begin{figure}
\begin{center}
\includegraphics[width=.45\textwidth]{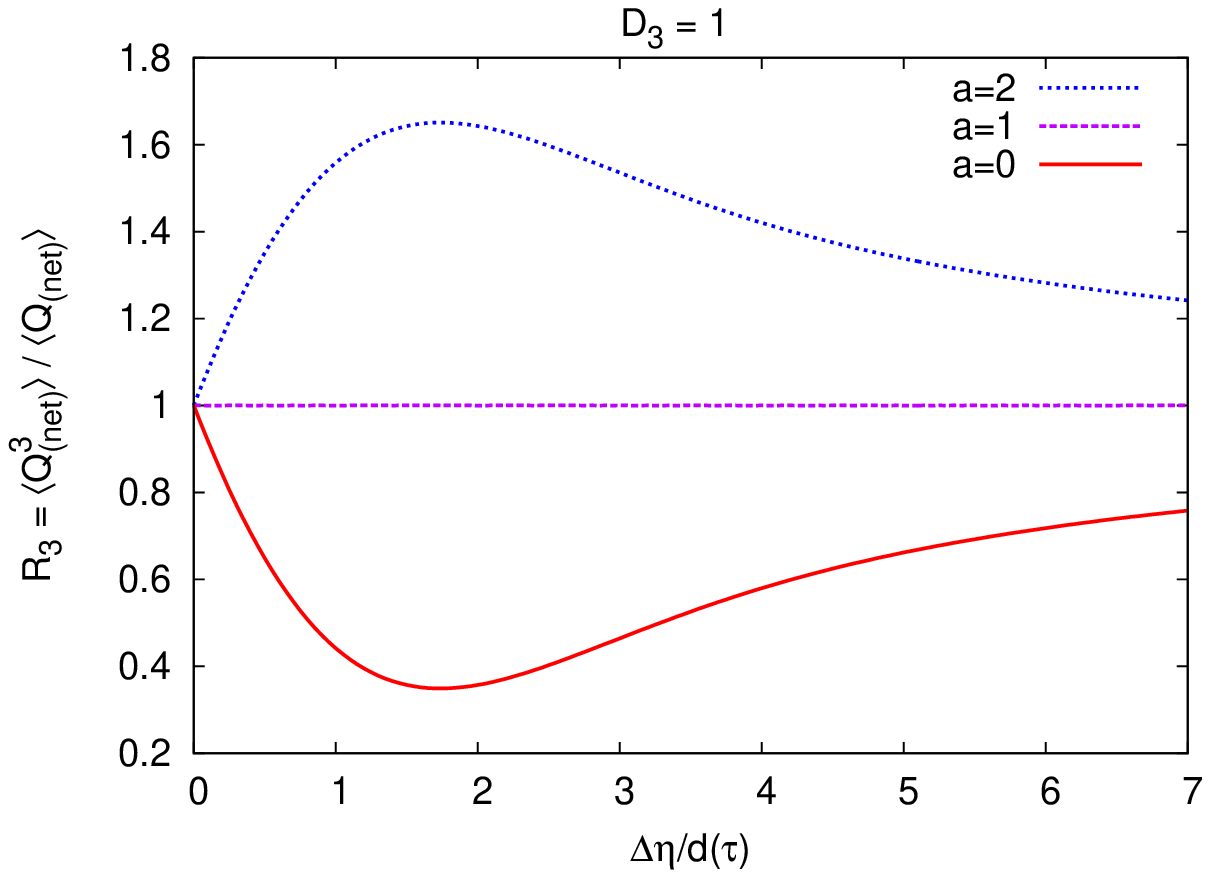}
\includegraphics[width=.45\textwidth]{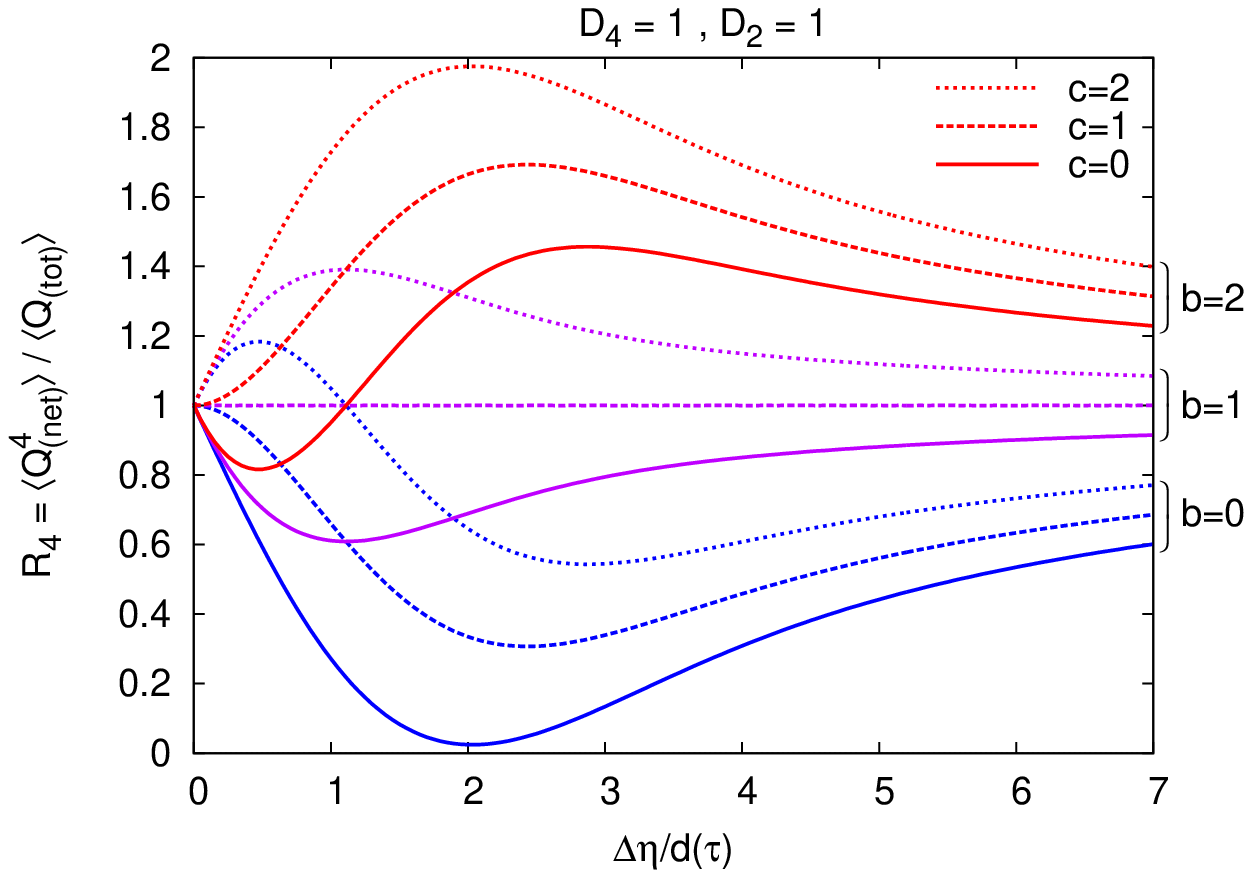}
\caption{
Dependences of the normalized non-Gaussian cumulants 
$R_3$ (left) and $R_4$ (right) on $\Delta\eta/d(\tau)$ 
with $D_2=D_3=D_4=1$.
}
\label{fig:R34_1}
\end{center}
\end{figure}

We first consider the case $D_2=D_3=D_4=1$, i.e. the
cumulants of the conserved charge in the initial condition 
is given by the Skellam distribution.
In Fig.~\ref{fig:R34_1}, we show the $\Delta\eta$
dependences of third and fourth order cumulants
with this initial condition.
In the figure, we vary the parameters $a$, $b$ and $c$,
which are not constrained by the conservation law.
The figure shows that the normalized cumulants
for a given $\Delta\eta$ can have a significant deviation 
from unity depending on the initial parameters although 
initial and equilibrated values of the cumulants are both unity.
Only when $a=b=c=1$, all normalized cumulants become 
unity irrespective of the value of $\Delta\eta$.
This result again tells us that the measurement of 
the cumulants with a fixed $\Delta\eta$ should not 
be compared with those in equilibrated medium.

\begin{figure}
\begin{center}
\includegraphics[width=.45\textwidth]{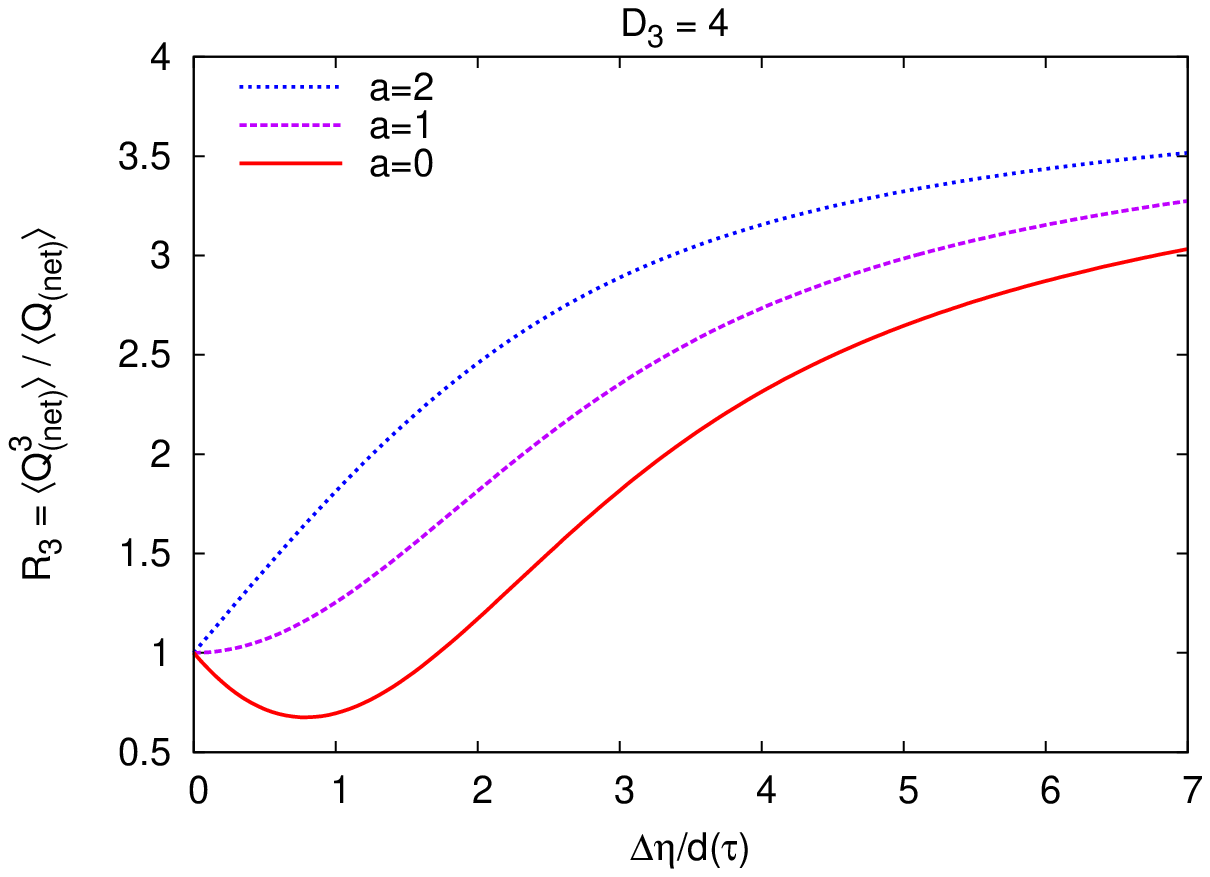}
\includegraphics[width=.45\textwidth]{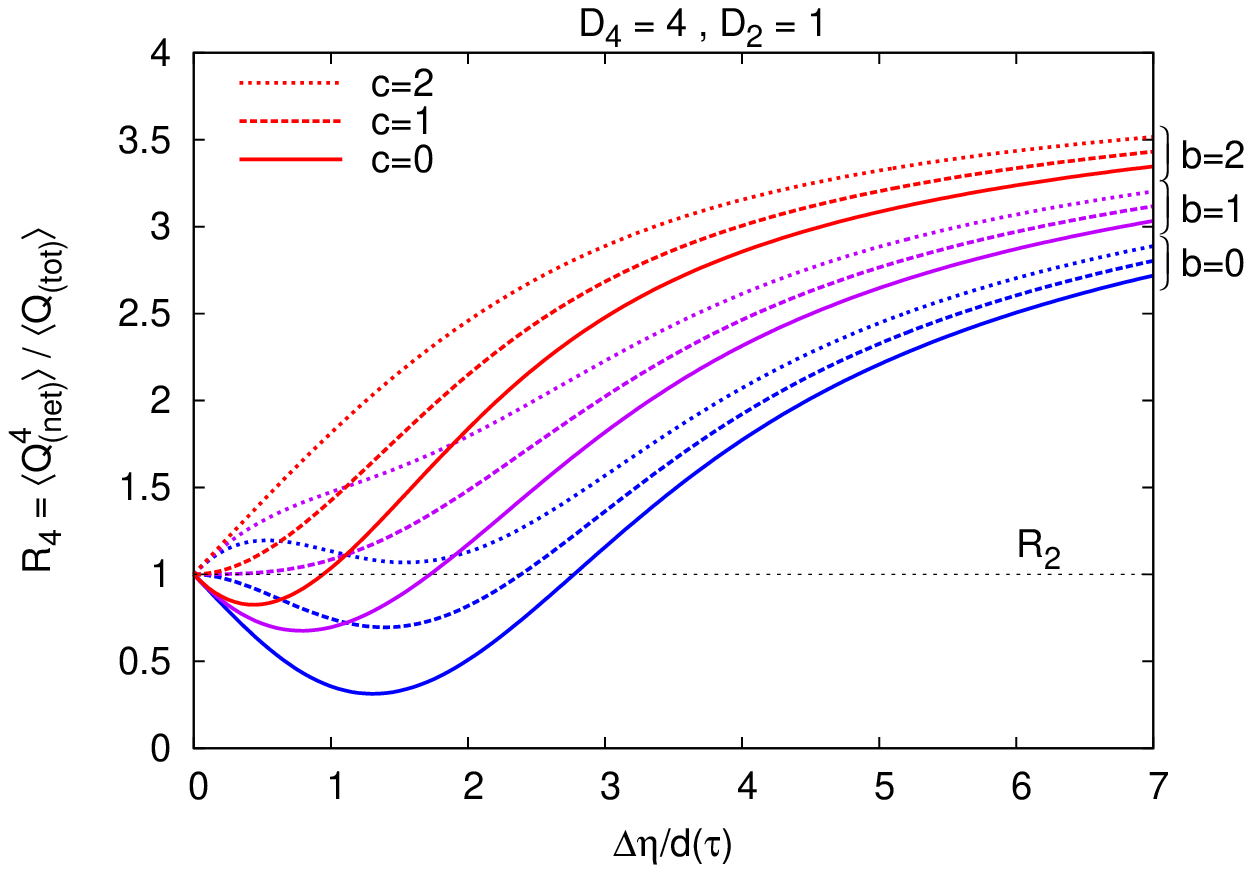}
\caption{
Dependences of the normalized non-Gaussian cumulants 
$R_3$ (left) and $R_4$ (right) on $\Delta\eta/d(\tau)$ 
with $D_3=4$ and $D_4=4$ with various 
initial parameters.
}
\label{fig:R34_4}
\end{center}
\end{figure}

Finally, we consider the $\Delta\eta$ dependence of 
the normalized cumulants $R_3$ and $R_4$ for relatively 
large $D_3$ and $D_4$.
Such a case would be realized when the medium near the 
critical point is formed in the time evolution of the hot 
medium and the critical enhancement of fluctuations 
are well developed \cite{Stephanov:2008qz}.
In Fig.~\ref{fig:R34_4} we show the $\Delta\eta$ dependences of 
$R_3$ and $R_4$ with $D_3=4$ and $D_4=4$, respectively, for 
several values of initial parameters.
The figure tells us the same conclusions as before:
(1) The values of $R_3$ and $R_4$ with a fixed $\Delta\eta$
can have significant deviation from the initial and 
equilibrated values, and (2) initial parameters can be 
constrained experimentally from the $\Delta\eta$ dependences 
of the normalized cumulants.

\section{Discussions and Summary}
\label{sec:summary}

In the previous section we have discussed the rapidity window 
$\Delta\eta$ dependence of the cumulants.
These results can be directly compared with the experimental 
results of the cumulants of net-baryon number and net-electric charge.
By this comparison, especially by the simultaneous use of 
the three normalized cumulants $R_2$, $R_3$ and $R_4$, 
parameters in the model and underlying physics 
on the thermal and transport properties of the hot medium 
can be constrained experimentally.
The discussion in the previous section also tells us the importance 
of the total understanding
of the $\Delta\eta$ dependences of the experimental results.
Because the value of the cumulants with a fixed $\Delta\eta$ 
can deviate significantly from the initial and final values, 
they should not be directly compared with theoretical studies
which are obtained in statistical mechanics assuming equilibration, 
such as lattice simulations.
The quantities which should be compared with the cumulants in 
statistical mechanics are $D_2$, $D_3$ and $D_4$, which can be 
deduced by the analysis of the $\Delta\eta$ dependence.

We note that the $\Delta\eta$ dependence of the Gaussian 
fluctuation encodes the same physical information as those 
accessible with the balance function 
\cite{Pratt:2012dz,Pratt:2014cza,Ling:2013ksb} 
of the corresponding particle species, because the former is 
obtained by the integral of the latter, and vice versa 
\cite{Bass:2000az,Jeon:2001ue,SAK}.
On the other hand, the $\Delta\eta$ dependences of non-Gaussian
cumulants are related to multi-particle correlation functions,
and provide us with a novel information on the 
diffusive process of conserved charges.

All the results in this study is obtained using the DME.
In this model it is assumed that the Brownian 
particles in the model do not interact with each other.
This assumption would be qualitatively justified 
when $Q_{\rm net}$ is identified to be the net-baryon number 
because of the following two reasons. 
First, baryons in the hadronic medium mainly 
interact with pions, and the baryon-baryon interactions 
scarcely take place \cite{KA2}.
Second, the success of the thermal model for the particle 
abundances indicates that the pair annihilation of baryons 
is well suppressed after chemical freezeout \cite{KA2}.
The baryons thus can be regarded as the Brownian particles 
floating in the pionic medium without interacting with one 
another, and the correlation between two baryons in the hadronic
medium will be well suppressed.
For net-electric charges, on the other hand, 
the interactions, especially the pair-annihilation, occur
frequently between hadrons having electric charge, especially pions.
The present model thus would not be suitable for the quantitative 
description of the diffusive process of the net-electric charge 
fluctuations.
It is desirable to extend the model to incorporate
the interaction between particles for more quantitative 
description of the net-electric charge fluctuations.
In the present study it is also assumed that the fluctuations
satisfy the locality condition at chemical freezeout time, 
which, however, might be violated in realistic collision events.
To incorporate such effects one has to modify the initial condition.
We left these extensions to future study and do not address
further in this study.

In this paper, we considered the time evolution and 
rapidity window, $\Delta\eta$, dependences of the cumulants of 
conserved charges observed in relativistic heavy ion collisions.
In order to describe the non-equilibrium time evolution 
of non-Gaussian cumulants we employed the diffusion master equation 
Eq.~(\ref{eq:DME}).
The analytic formula for the time evolution of cumulants 
with arbitrary initial conditions are obtained.
Using these results, we discussed the $\Delta\eta$ dependence of 
cumulants of conserved charges observed in heavy ion collisions.
Various suggestions have been made to utilize the $\Delta\eta$ 
dependence of the non-Gaussian cumulants for the understanding 
of the thermal and transport properties of the hot medium.

The author thanks Masayuki~Asakawa, Hirosato~Ono and 
Miki~Sakaida for valuable discussions.
He also thanks Shinichi~Esumi, Volker~Koch, Hiroshi~Masui and 
Krzysztof~Redlich for useful conversations and encouragements.
This work was supported by JSPS KAKENHI (Grant Number 25800148).

\appendix

\section{Superposition of cumulants}
\label{app:superposition}

In this appendix, we consider cumulants of a probability 
distribution function which is given by a superposition 
of probability distribution functions.
Let us consider a probability distribution function $P(x)$
for an integer stochastic variable $x$, and assume that 
$P(x)$ consists of the superposition of sub-probabilities as 
\begin{align}
P(x) = \sum_N F(N) P_N(x),
\label{eq:PFPN}
\end{align}
where $P_N(x)$ are sub-probabilities 
labeled by integer $N$. 
Each sub-probability is summed with a weight
$F(N)$ satisfying $\sum_N F(N)=1$ , which is also regarded 
as a probability.
The purpose of this appendix is to represent 
the cumulants of $P(x)$ using those of $P_N(x)$
and $F(N)$.
Although we write down the results explicitly 
up to fourth order in this manuscript, 
the result can be extended to higher orders straightforwardly.
A similar result given in this appendix is presented 
and used in Ref.~\cite{KA2} to relate the net-proton and 
net-baryon number cumulants in the final state in 
relativistic heavy ion collisions.

We start from the cumulant generating function of 
Eq.~(\ref{eq:PFPN}),
\begin{align}
K(\theta) 
&= \log \sum_x e^{ \theta x } P(x)
= \log \sum_N F(N) \sum_x e^{ \theta x } P_N(x)
\\
&= \log \sum_N F(N) \sum_x e^{ K_N(\theta) }
\label{eq:KNdef}
\end{align}
where $K_N(\theta) = \log \sum_x e^{ \theta x } P_N(x)$ is 
the cumulant generating function for $P_N(x)$.
Using the cumulant expansion, 
Eq.~(\ref{eq:KNdef}) is written as
\begin{align}
K(\theta)
=& \sum_m \frac1{m!} \sum_F [ K_N(\theta) ]_c^m
\\
=& \sum_F K_N(\theta) + \frac12 \sum_F ( \delta K_N(\theta) )^2
+ \frac1{3!} \sum_F ( \delta K_N(\theta) )^3
\nonumber
\\
&+ \frac1{4!} \sum_F [ K_N(\theta) ]_c^4
+ \cdots,
\label{eq:K1234}
\end{align}
where $\sum_F$ is the shorthand notation of $\sum_N F(N)$,
and $\sum_F [ K_N(\theta) ]_c^m$ is the $m$th-order cumulant
of $K_N$ for the sum over $F$, 
whose explicit forms up to fourth order are given in 
the far right hand side with 
\begin{align}
\sum_F (\delta K_N(\theta))^n 
&= \sum_F \left( K_N(\theta) - \sum_F K_N(\theta) \right)^n ,
\\
\sum_F [ K_N(\theta)]_c^4
&= \sum_F (\delta K_N(\theta))^4
- 3 \left( \sum_F (\delta K_N(\theta))^2 \right)^2 .
\end{align}

Cumulants of $P(x)$ are given by derivatives of $K(\theta)$ as 
\begin{align}
\langle x^n \rangle_c 
= \frac{ \partial^n }{ \partial \theta^n } K(0) \equiv K^{(n)}.
\label{eq:<x^n>c}
\end{align}
All cumulants can be obtained with Eqs.~(\ref{eq:<x^n>c}) 
and (\ref{eq:K1234}).
In order to calculate the cumulants explicitly,
we first note that the normalization condition 
$\sum_x P_N(x)=1$ yields $K_N(\theta)=0$. 
From this property, it is immediately concluded that 
all $K_N(\theta)$ in a term in the far right hand 
side of Eq.~(\ref{eq:K1234}) must receive at least 
one differentiation so that the term gives nonzero 
contribution to Eq.~(\ref{eq:<x^n>c}).
This means that the $m$th order term in 
Eq.~(\ref{eq:K1234}) can affect Eq.~(\ref{eq:<x^n>c})
only if $m\le n$.
Keeping this rule in mind, derivatives of 
Eq.~(\ref{eq:K1234}) with $\theta=0$ is given by
\begin{align}
K^{(1)}
=& \sum_F K_N^{(1)},
\label{eq:K1KN}
\\
K^{(2)}
=& \sum_F K_N^{(2)} + \sum_F ( \delta K_N^{(1)} )^2,
\label{eq:K2KN}
\\
K^{(3)}
=& \sum_F K_N^{(3)} + 3 \sum_F \delta K_N^{(1)} \delta K_N^{(2)} 
+ \sum_F ( \delta K_N^{(1)} )^3,
\label{eq:K3KN}
\\
K^{(4)}
=& \sum_F K_N^{(4)} + 4 \sum_F \delta K_N^{(1)} \delta K_N^{(3)} 
+ 3 \sum_F ( \delta K_N^{(2)} )^2
+ 6 \sum_F ( \delta K_N^{(1)} )^2 \delta K_N^{(2)}
\nonumber
\\ &
+ \sum_F ( \delta K_N^{(1)} )_c^4,
\label{eq:K4KN}
\end{align}
with $K_N^{(n)}=\partial^n K_N(0) / \partial \theta^n$
being the cumulants of sub-probabilities $P_N(x)$.
Equations~(\ref{eq:K1KN}) - (\ref{eq:K4KN}) relate 
the cumulants $K^{(n)}$ with $K_N^{(n)}$.

The above relations are further simplified when 
the cumulants of $P_N(x)$ are at most linear with
respect to $N$, i.e.
\begin{align}
K_N^{(n)}= N \xi_{(n)} + \zeta_{(n)} ,
\label{eq:K_Nxi}
\end{align}
where $\xi_{(n)}$ and $\zeta_{(n)}$ are constants 
which do not depend on $N$.
Substituting Eq.~(\ref{eq:K_Nxi}) in 
Eqs.~(\ref{eq:K1KN}) - (\ref{eq:K4KN})
one obtains
\begin{align}
K^{(1)} 
=& \zeta_{(1)} + \xi_{(1)} \langle N \rangle_F ,
\label{eq:K1NF}
\\
K^{(2)} 
=& \zeta_{(2)} + \xi_{(2)} \langle N \rangle_F 
+ {\xi_{(1)}}^2 \langle \delta N^2 \rangle_F ,
\label{eq:K2NF}
\\
K^{(3)} 
=& \zeta_{(3)} + \xi_{(3)} \langle N \rangle_F 
+ 3 \xi_{(1)} \xi_{(2)} \langle \delta N^2 \rangle_F 
+ {\xi_{(1)}}^3 \langle \delta N^3 \rangle_F ,
\label{eq:K3NF}
\\
K^{(4)} 
=& \zeta_{(4)} + \xi_{(4)} \langle N \rangle_F 
+ ( 4 \xi_{(1)} \xi_{(3)} + 3 {\xi_{(2)}}^2 ) \langle \delta N^2 \rangle_F 
+ 6 {\xi_{(1)}}^2 \xi_{(2)} \langle \delta N^3 \rangle_F
\nonumber \\ &
+ {\xi_{(1)}}^4 \langle \delta N^4 \rangle_{c,F} ,
\label{eq:K4NF}
\end{align}
where $\langle O(N) \rangle_F = \sum_N O(N) F(N)$ 
denotes the average over $F(N)$; these averages in 
the above formulas represent cumulants of the probability $F(N)$.

These results are easily generalized to 
the case where the sub-probabilities are labeled
by multi-dimensional integers,
\begin{align}
P(x) = \sum_{\bm N} F(\bm{N}) P_{\bm N}(x),
\label{eq:PFPNM}
\end{align}
with $\bm{N}=(N_1,N_2,\cdots,N_m)$ representing
a set of $m$ integers.
When the cumulants of $P_{\bm N}(x)$ are linear in $\bm{N}$, i.e.,
\begin{align}
K_{\bm N}^{(n)} = \bm{N}\cdot \bm{\xi}_{(n)} + \zeta_{(n)},
\end{align}
a similar manipulation as before yields,
\begin{align}
K^{(1)} 
=& \zeta_{(1)} + \langle \bm{\xi}_{(1)} \cdot \bm{N} \rangle_F ,
\label{eq:K1NFM}
\\
K^{(2)} 
=& \zeta_{(2)} + \langle \bm{\xi}_{(2)} \cdot \bm{N} \rangle_F
+ \langle ( \bm{\xi}_{(1)} \cdot \delta \bm{N})^2 \rangle_F ,
\label{eq:K2NFM}
\\
K^{(3)} 
=& \zeta_{(3)} + \langle \bm{\xi}_{(3)}\cdot\bm{N} \rangle_F 
+ 3 \langle (\bm{\xi}_{(1)}\cdot \delta\bm{N})
(\bm{\xi}_{(2)}\cdot \delta\bm{N}) \rangle_F 
+ \langle (\bm{\xi}_{(1)}\cdot \delta\bm{N})^3 \rangle_F ,
\label{eq:K3NFM}
\\
K^{(4)} 
=& \zeta_{(4)} + \langle \bm{\xi}_{(4)}\cdot \bm{N} \rangle_F 
+ 4 \langle (\bm{\xi}_{(1)}\cdot \delta\bm{N})
(\bm{\xi}_{(3)}\cdot \delta\bm{N}) \rangle_F 
+ 3 \langle (\bm{\xi}_{(2)}\cdot \delta\bm{N})^2 \rangle_F 
\nonumber \\ &
+ 6 \langle (\bm{\xi}_{(1)}\cdot \delta\bm{N})^2
(\bm{\xi}_{(2)}\cdot \delta\bm{N}) \rangle_F 
+ \langle (\bm{\xi}_{(1)}\cdot \delta\bm{N})^4 \rangle_{c,F}.
\label{eq:K4NFM}
\end{align}

\section{Chemical Reaction}
\label{app:chemical}

In this appendix, we consider a simple chemical reaction
between two species of particles X and A \cite{Gardiner},
\begin{align}
{\rm X} \rightleftharpoons {\rm A}.
\end{align}
It is assumed that a particle X (A) becomes A (X) with 
a probability $k_1$ ($k_2$) per unit time.
We denote the number of each particle in the system as $x$ and $a$.
For simplicity, it is further assumed that $a$ is 
sufficiently large compared with $x$ so that the number 
$a$ can be regarded fixed.
The time evolution of the probability distribution $P(x,t)$ 
is then described by the master equation \cite{Gardiner}
\begin{align}
\partial_t P(x,t) 
= k_2 a P( x-1,t ) + k_1 ( x+1 ) P( x+1,t )
- ( k_1 x + k_2 a ) P( x,t ).
\label{eq:XA_master}
\end{align}
In the following, we present the solution of this 
master equation for arbitrary initial conditions.

\subsection{Fixed initial condition}

Let us first solve Eq.~(\ref{eq:XA_master}) with 
a fixed initial condition
\begin{align}
P( x,0 ) = \delta_{x,N},
\label{eq:XA_init}
\end{align}
i.e. the number $x$ is fixed to $N$ without 
fluctuations at $t=0$.

In order to solve Eq.~(\ref{eq:XA_master}), 
we use factorial generating function,
\begin{align}
G_{\rm f}( s,t ) 
= \sum_x s^x P( x,t ).
\label{eq:XA_G}
\end{align}
Substituting Eq.~(\ref{eq:XA_G}) in Eq.~(\ref{eq:XA_master}),
one obtains 
\begin{align}
\partial_t G_{\rm f}( s,t ) 
= k_2 a ( s-1 ) G_{\rm f}( s,t ) 
- k_1 ( s-1 ) \partial_s G_{\rm f}( s,t ).
\label{eq:XA_Gt}
\end{align}
Equation~(\ref{eq:XA_Gt}) can be 
solved by using the method of characteristics, 
which yields
\begin{align}
G_{\rm f}( s,t ) = F( (s-1)e^{-k_1 t} ) e^{(s-1)N_{\rm eq}}
\end{align}
with an arbitrary function $F$ and the average 
number of $x$ in equilibrium, $N_{\rm eq} = k_2 a / k_1$.

The function $F$ is determined by specifying the 
initial condition. 
The generating function corresponding to the initial
condition Eq.~(\ref{eq:XA_init}) is 
$G_{\rm f}( s,0 ) = s^N$. This gives
\begin{align}
F(r) = (r+1)^N e^{rN_{\rm eq}}.
\end{align}
The solution of Eq.~(\ref{eq:XA_Gt}) with the 
initial condition Eq.~(\ref{eq:XA_init}) thus is 
\begin{align}
G_{\rm f}( s,t ) 
= ( 1 + (s-1) e^{-k_1 t} )^N e^{ N_{\rm eq} (s-1)(1-e^{-k_1 t}) }.
\end{align}

The moment and cumulant generating functions, $G(\theta,t)$ 
and $K(\theta,t)$, respectively, are given by
\begin{align}
G( \theta,t ) &= G_{\rm f}( e^\theta,t ) ,
\\
K( \theta,t ) &= \log G( \theta,t ) .
\end{align}
Derivatives of $K(\theta,t)$ define cumulants; 
$\langle x(t)^n \rangle_c = \partial^n K(0,t) / \partial \theta^n$.
Using 
\begin{align}
\frac{ \partial }{ \partial \theta } G( \theta,t )
&= \frac{ \partial s }{ \partial \theta } 
\frac{ \partial }{ \partial s } G_f( s,t )
= e^\theta \frac{ \partial }{ \partial s } G_f( s,t ),
\\
\frac{ \partial^2 }{ \partial \theta^2 } G( \theta,t )
&= e^\theta \frac{ \partial }{ \partial s } G_f( s,t )
+ e^{2\theta} \frac{ \partial^2 }{ \partial s^2 } G_f( s,t ),
\end{align}
and so forth, one obtains
\begin{align}
\langle x(t) \rangle_c 
&= N_{\rm eq} ( 1 - e^{ -k_1 t } ) + N e^{ -k_1 t } ,
\label{eq:XA1}
\\
\langle x(t)^2 \rangle_c 
&= N_{\rm eq} ( 1 - e^{ -k_1 t } ) + N ( e^{ -k_1 t } - e^{ -2k_1 t } ) ,
\label{eq:XA2}
\\
\langle x(t)^3 \rangle_c 
&= N_{\rm eq} ( 1 - e^{ -k_1 t } ) + N ( e^{ -k_1 t } -3 e^{ -2k_1 t } +2 e^{ -3k_1 t } ),
\label{eq:XA3}
\\
\langle x(t)^4 \rangle_c 
&= N_{\rm eq} ( 1 - e^{ -k_1 t } )
+ N ( e^{ -k_1 t } -7 e^{ -2k_1 t } +12 e^{ -3k_1 t } -6 e^{ -4k_1 t } ) .
\label{eq:XA4}
\end{align}

\subsection{General initial condition}

In order to obtain the $t$ dependence of the cumulants for 
general initial conditions, we make use of the result in 
Appendix~\ref{app:superposition}.
Since Eq.~(\ref{eq:XA_master}) is a linear differential
equation, the solution of $P(x,t)$ for an initial
condition $P(x,0) = F(x)$ is given by 
\begin{align}
P(x,t) = \sum_N F(N) P_N(x,t),
\end{align}
where $P_N(x,t)$ is the solution for the fixed initial
condition, Eq.~(\ref{eq:XA_init}).
We further remark that the cumulants 
Eqs.~(\ref{eq:XA1}) - (\ref{eq:XA4}) are at most linear 
in $N$ and fulfill the form in Eq.~(\ref{eq:K_Nxi})
with 
\begin{align}
\zeta_{(n)} = N_{\rm eq} ( 1 - e^{ -k_1 t } ) , \quad
\xi_{(1)} = e^{ -k_1 t } , 
\quad
\xi_{(2)} = e^{ -k_1 t } - e^{ -2k_1 t } ,
\label{eq:zeta,xi}
\end{align}
and etc.
The cumulants for arbitrary initial conditions thus are 
obtained by substituting Eqs.~(\ref{eq:zeta,xi}) 
in Eqs.~(\ref{eq:K1NF}) - (\ref{eq:K4NF}).The results up to fourth order are
\begin{align}
\langle x(t) \rangle_c 
=& N_{\rm eq} ( 1 - e^{ -k_1 t } ) + \langle N \rangle_0 e^{ -k_1 t },
\label{eq:XA1G}
\\
\langle x(t)^2 \rangle_c 
=& N_{\rm eq} ( 1 - e^{ -k_1 t } )
+ \langle N \rangle_0 e^{ -k_1 t }
+ ( \langle \delta N^2 \rangle_0 - \langle N \rangle_0 ) e^{ -2k_1 t },
\label{eq:XA2G}
\\
\langle x(t)^3 \rangle_c 
=& N_{\rm eq} ( 1 - e^{ -k_1 t } )
+ \langle N \rangle_0 e^{ -k_1 t }
+ 3 ( \langle \delta N^2 \rangle_0 - \langle N \rangle_0 ) e^{ -2k_1 t }
\nonumber \\ &
+ ( \langle \delta N^3 \rangle_0 - 3 \langle \delta N^2 \rangle_0 
+ 2 \langle N \rangle_0 ) e^{ -3k_1 t } ,
\label{eq:XA3G}
\\
\langle x(t)^4 \rangle_c 
=& N_{\rm eq} ( 1 - e^{ -k_1 t } )
+ \langle N \rangle_0 e^{ -k_1 t }
+ 7 ( \langle \delta N^2 \rangle_0 - \langle N \rangle_0 ) e^{ -2k_1 t }
\nonumber \\ &
+ 6 ( \langle \delta N^3 \rangle_0 - 3 \langle \delta N^2 \rangle_0 
+ 2 \langle N \rangle_0 ) e^{ -3k_1 t }
\nonumber \\ &
+ ( \langle \delta N^4 \rangle_{c,0} - 6 \langle \delta N^3 \rangle_0 
+ 11 \langle \delta N^2 \rangle_0 - 6 \langle N \rangle_0 ) e^{ -4k_1 t },
\label{eq:XA4G}
\end{align}
where $\langle O(N) \rangle_0$ denotes the average in 
the initial condition.

From Eqs.~(\ref{eq:XA1G}) - (\ref{eq:XA4G}), one 
finds that in the large $t$ limit, $k_1 t\to\infty$,
all cumulants converge a same value,
\begin{align}
\langle x(t)^n \rangle_c = N_{\rm eq} ,
\end{align}
which means that the distribution approaches the Poissonian
in the large $t$ limit irrespective of the initial condition.

\subsection{Some examples of time evolution}

It is instructive to see the time evolution of 
$\langle x(t)^n \rangle_c$ 
with some specific initial conditions.

\subsubsection{Poisson distribution}

First, let us consider the initial condition that 
$P(x,t)$ is the Poisson distribution at $t=0$. 
In this case, the initial condition satisfies
\begin{align}
\langle N \rangle_0 = \langle \delta N^2 \rangle_0
= \langle \delta N^3 \rangle_0 = \langle \delta N^4 \rangle_{c,0}.
\label{eq:N0poisson}
\end{align}
Substituting Eq.~(\ref{eq:N0poisson}) 
in Eq.~(\ref{eq:XA1G}) - (\ref{eq:XA4G}), all terms 
in the right-hand side vanish except for the first two terms 
and one finds that the all cumulants have the same $t$ 
dependence 
\begin{align}
\langle x(t)^n \rangle_c = 
N_{\rm eq} ( 1 - e^{ -k_1 t } )
+ \langle N \rangle_0 e^{ -k_1 t } .
\end{align}
This result shows that the distribution $P(x,t)$ starting
from the Poissonian stays Poissonian for all $t$ \cite{Gardiner}, 
while the average of the distribution shifts from 
$\langle N \rangle_0$ at $t=0$ to $N_{\rm eq}$
for $t\to\infty$.

\subsubsection{$\langle N \rangle_0 = N_{\rm eq}$}

\begin{figure}
\begin{center}
\includegraphics[width=.49\textwidth]{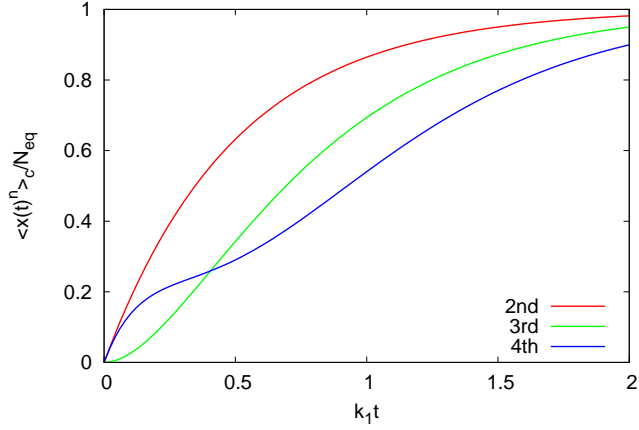}
\caption{
Dependences of the cumulants $\langle x(t)^n \rangle_c$
on $t$ with fixed initial condition 
Eq.~(\ref{eq:chem_fixed}).
}
\label{fig:chem}
\end{center}
\end{figure}

Next, we consider the initial condition
\begin{align}
P(x,0) = \delta_{x,N_{\rm eq}},
\label{eq:chem_fixed}
\end{align}
i.e. the particle number is fixed to the equilibrated 
value $N_{\rm eq}$ 
without fluctuation.
In this case, one obtains
\begin{align}
\langle x(t) \rangle_c 
=& N_{\rm eq} ,
\\
\langle x(t)^2 \rangle_c 
=& N_{\rm eq} ( 1 - e^{ -2k_1 t } ),
\\
\langle x(t)^3 \rangle_c 
=& N_{\rm eq} 
( 1 - 3 e^{ -2k_1 t } + 2 e^{ -3k_1 t } ),
\\
\langle x(t)^4 \rangle_c 
=& N_{\rm eq} ( 1 - 7 e^{ -2k_1 t } + 12 e^{ -3k_1 t } - 6 e^{ -4k_1 t } ).
\end{align}
In Fig.~\ref{fig:chem}, we show the $t$ dependence 
of second to fourth order cumulants for the initial condition 
Eq.~(\ref{eq:chem_fixed}) up to fourth order.
The figure shows that all cumulants approach $N_{\rm eq}$ 
as $t$ increases.
One also finds that the growth of higher order cumulants is slower.
This behavior is understood as follows.
First, the higher order cumulants are more sensitive to 
the tails, i.e. far side from the average value, of the distribution.
Second, propagation of the probability
distribution to far side from the $N_{\rm eq}$ should take longer time.
Thus, the approach of higher order cumulants to the 
equilibrated value is much slower than the lower order
ones.

\section{$I_X(\zeta)$ and $F_X$ }
\label{app:I_X}

In this appendix, we summarize property of 
$I_X(\zeta)$ defined in Eq.~(\ref{eq:I_X})
\begin{align}
I_X(\zeta) =& \int_{-1/2}^{1/2} d\xi \int \frac{dp}{2\pi}
e^{-X^2 p^2/2} e^{ip(\xi+\zeta)} ,
\label{eq:I_X:app}
\end{align}
and $F_n(X)$ defined in Eq.~(\ref{eq:F_n}).

By first integrating out $p$ in 
Eq.~(\ref{eq:I_X:app}) one obtains
\begin{align}
I_X(\zeta) =& \int_{-1/2}^{1/2} d\xi 
\frac1{\sqrt{2\pi}X} e^{-(\xi+\zeta)^2/2X^2}
\nonumber \\
=& \frac12 \left(
{\rm erf} \frac{ \zeta+1/2 }{\sqrt{2}X} 
- {\rm erf} \frac{ \zeta-1/2 }{\sqrt{2}X} \right).
\end{align}

For $X=0$, $I_X(\zeta)$ satisfies 
\begin{align}
I_0(\zeta) = \int_{-1/2}^{1/2} d\xi \int \frac{dk}{2\pi} e^{ik(\xi+\zeta)}
= \int_{-1/2}^{1/2} d\xi \delta(\xi-\zeta)
= \theta( 1-\zeta^2 ),
\end{align}
where $\theta(x)$ is the step function.
From this result, one obtains
\begin{align}
[I_0(\zeta)]^n = \theta( 1-\zeta^2 ),
\label{eq:I_0(z)}
\end{align}
for arbitrary $n$.

\begin{figure}
\begin{center}
\includegraphics[width=.49\textwidth]{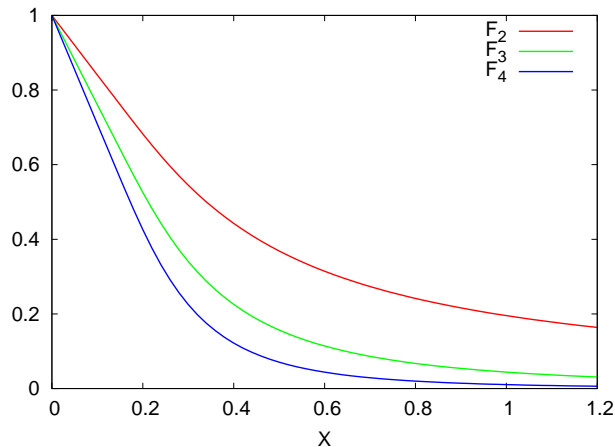}
\caption{
Functions $F_n(X)$ with $n=2,3$ and $4$.
}
\label{fig:F234T}
\end{center}
\end{figure}

In Sec.~\ref{sec:DME}, we define
\begin{align}
F_n(X) = \int_{-\infty}^{\infty} dz [I_X(z/\Delta)]^n.
\end{align}
From Eq.~(\ref{eq:I_0(z)}), one easily obtains
\begin{align}
F_n(0) = 1
\end{align}
for any $n$.
The integral for $n=2$ can be performed analytically
and one obtains
\begin{align}
F_2(X) = {\rm erf} \frac1{2X}
-\frac{2X}{\sqrt{\pi}} (1-e^{-1/4X^2}) .
\end{align}
For $n=1$, $F_1(X)$ is a constant
\begin{align}
F_1(X) = 1.
\end{align}
For $n\ge2$, $F_n(X)$ are monotonically 
decreasing functions of $X$ with $F_n(0)=1$ and 
$\lim_{X\to\infty} F_n(X)=0$.
In Fig.~\ref{fig:F234T}, we show the $X$ dependences of 
$F_n(X)$ for $n=2,3$ and $4$.

\end{document}